# Crossover between intrinsic and temperature-assisted regimes in spin-orbit torque switching of antiferromagnetic order

Takumi Matsuo[1,2,†], Tomoya Higo[1,3,4†,*], Daisuke Nishio-Hamane[4], Takuya Matsuda[1,5], Ryota Uesugi[1,4], Hanshen Tsai[1], Kouta Kondou[6,7], Shinji Miwa[4], Yoshichika Otani[4,6], and Satoru Nakatsuji[1,2,4,8,9*]

[1]*Department of Physics, The University of Tokyo, Bunkyo-ku, Tokyo 113-0033, Japan*

[2]*Department of Physics and Astronomy, The Johns Hopkins University, Baltimore, Maryland 21218, USA*

[3]*PRESTO, Japan Science and Technology Agency, Kawaguchi, Saitama 332-0012, Japan*

[4]*Institute for Solid State Physics, The University of Tokyo, Kashiwa, Chiba 277-8581, Japan.*

[5]*Graduate School of Engineering Science, Osaka University, Toyonaka, Osaka 560-8531, Japan*

[6]*Center for Emergent Matter Science, RIKEN, Wako, Saitama 351-0198, Japan*

[7]*Institute for Open and Transdisciplinary Research Initiatives, Osaka University, Toyonaka, Osaka 560-8531, Japan*

[8]*Trans-scale Quantum Science Institute, The University of Tokyo, Bunkyo-ku, Tokyo 113-0033, Japan*

[9]*Canadian Institute for Advanced Research, Toronto, Ontario M5G 1Z7, Canada*

†These authors equally contributed to this work

*Corresponding Author: tomoyahigo@g.ecc.u-tokyo.ac.jp, satoru@g.ecc.u-tokyo.ac.jp

## Abstract

**Intensive studies have been made on antiferromagnets as candidate materials for next generation memory bits due to their ultrafast dynamics reaching picosecond time scales. Recent demonstrations of electrical bidirectional switching of antiferromagnetic states have attracted significant attention. However, under the presence of significant Joule heating that destabilizes the magnetic order, the timescales associated with the switching can be limited to nanoseconds or longer. Here, we present the observation of a crossover in the switching behavior of the chiral antiferromagnet $Mn_3Sn$ by tuning the magnetic layer thickness. While Joule heating interferes with switching in thicker devices, we find clear signatures of an**

**intrinsic spin-orbit torque mechanism as the thickness is reduced, avoiding the heating effect. The suppression of heating enables switching without significant attenuation of the readout signal using pulses shorter than those required by temperature-assisted mechanisms. The crossover into the spin-orbit torque switching behavior clarifies the potential for achieving ultrafast switching as expected from the picosecond spin dynamics of antiferromagnets. Our results lay the groundwork for designing antiferromagnetic memory devices that can operate at ultrafast timescales.**

## Introduction

The ability to electrically manipulate and detect the orientation of magnetization in ferromagnetic order has led to the advent of non-volatile magnetic memory[1–3]. In particular, the realization of perpendicular magnetic recording technology has enabled stable and reliable operations in nanoscale ferromagnetic bits[3–5]. On the other hand, meeting the increasing demand for information storage and processing requires memory bits that enable higher packing densities and faster data manipulation.

Replacing ferromagnets with antiferromagnets in memory bits is a promising route to meet these benchmarks. The absence of stray magnetic field in antiferromagnets due to lack of magnetization allows for high integration density, while the picosecond order spin dynamics of antiferromagnets is promising in achieving processing time scales two to three orders of magnitude faster than the nanosecond operations accessible in conventional ferromagnetic devices[6–12]. However, due to their vanishingly small magnetization, the small responses of typical antiferromagnets have hindered their application role. This bottleneck was overcome by the discovery of antiferromagnets with unique magnetic ordering that breaks global time-reversal symmetry, showing large responses beyond the conventional magnetization scaling behavior[13–22]. In these antiferromagnets, it has been demonstrated that their magnetic ordering can be manipulated using spin-orbit torque (SOT)[23–33] with a protocol similar to that used for ferromagnets[4,34,35], and that perpendicular magnetic recording is also possible[28], making them promising candidates for materials that can realize next-generation memory storage.

When using ferromagnets, manufacturing devices with perpendicular magnetization requires careful consideration of device thickness. The large shape anisotropy in ferromagnetic films prevents the magnetization from pointing in the out-of-plane direction. To counteract this demagnetization effect, a sufficiently large perpendicular magnetic anisotropy (PMA) is required. While materials with strong magnetocrystalline anisotropy are being developed, the use of interfacial perpendicular magnetic anisotropy has been the mainstream approach[36]. This interfacial anisotropy scales inversely with the thickness of the magnetic layer. As a result, the thickness range allowed for perpendicular ferromagnetic devices is typically limited to up to several nanometers[3,4,34,35]. Antiferromagnets, on the other hand, can retain perpendicular magnetic order across a much wider range of thicknesses as the demagnetization energy is negligibly small owing to their small magnetization[37], and PMA can be achieved simply under the presence of a small magnetic anisotropy regardless of thickness.

In perpendicularly magnetized ferromagnetic layers, the device thickness is also constrained by the magnetic exchange length $t_{\mathrm{ex}}$. $t_{\mathrm{ex}}$ is defined as $\sqrt{\frac{A}{K}}$, where $A$ is the exchange constant and $K$ is the magnetic anisotropy constant per volume[38]. It can be interpreted as the size below which a magnet behaves approximately as a single magnetic domain[39]. In the context of magnetic switching, $t_{\mathrm{ex}}$ corresponds to the switchable magnetic layer thickness through spin current injection. In perpendicular ferromagnetic devices, a large $K$ is required, and even if a thicker ferromagnetic layer could be used, $t_{\mathrm{ex}}$ would be small, limiting the functional thickness of the device. Indeed, narrow domain walls of less than 10 nm have been reported in magnets with significant PMA[40–42], implying correspondingly small values of $t_{\mathrm{ex}}$. In contrast, for perpendicular antiferromagnetic devices, a large $K$ is not required, so $t_{\mathrm{ex}}$ becomes much larger than in ferromagnets with PMA, resulting in a large switchable thickness for the magnetic layer.

In fact, manipulation of antiferromagnetic order via SOT has been reported in devices with thicknesses from several to hundreds of nanometers[24,25,29,30,32,43]. The current densities required for bidirectional SOT switching ($j_{\mathrm{C}}$) in antiferromagnets with these thicknesses are on the order of 1-10 MA/cm$^2$ [23,24,28–30,32,43], comparable to or slightly smaller than the values in typical ferromagnetic devices[5,44].This demonstrates the viability of antiferromagnetic devices across a wide range of thicknesses, which could be useful in optimizing properties such as thermal stability. However, the flexibility in film thickness offered by antiferromagnetic devices, which is not available for ferromagnetic devices (Fig. 1**a**), introduces an important consideration when examining the detailed mechanisms of SOT switching.

The noncollinear antiferromagnet $Mn_3Sn$ has been frequently used as a model material for studying the functionalities necessary for designing antiferromagnetic memory[24–33,43,45–51]. Its large responses, such as the anomalous Hall effect (AHE)[13,15,52], anomalous Nernst effect[37,53,54], and the magneto-optical Kerr effect[55] enable the easy detection of its antiferromagnetic order despite its very weak magnetic moment of 3 m$\mu_B$/Mn[52,56–58]. These responses arise from the breaking of the time-reversal symmetry (TRS) due to the ferroic ($Q$ = 0) order of the cluster magnetic octupole[52], a higher-rank magnetic multipole consisting of six Mn moments in a unit cluster, hosted by the three-sublattice chiral antiferromagnetic structure of $Mn_3Sn$[13,56,57,59–62] (Fig. 1**b**). The Berry curvature distribution in momentum space, arising from the Weyl points in the band structure[14,17].

The importance of $Mn_3Sn$ as a functional antiferromagnet is underlined by the capability to bidirectionally and deterministically switch its magnetic order through SOT[24–33,51,63]. To achieve SOT switching in $Mn_3Sn$, a current pulse is applied to a $Mn_3Sn$/heavy metal (HM) bilayer device. In this device configuration, the spin current generated by the spin Hall effect (SHE) in the HM layer is injected into the $Mn_3Sn$ layer. (Notably, this setup is identical to that employed for conventional ferromagnets[4,34,35].) The injected spin torque instantly switches the chiral antiferromagnetic order in $Mn_3Sn$, which can be detected by a change in the sign of the AHE. The demonstration of bidirectional and perpendicular switching makes $Mn_3Sn$ the ideal candidate material for next-generation magnetic memory that enables picosecond data processing and has motivated extensive studies on the electrical manipulation and detection of its chiral antiferromagnetic order[27,33,43,45–51,64].

In the SOT switching protocol described above, the injection of a write current pulse inevitably heats the magnetic layer to a certain temperature $T_{\mathrm{switch}}$ (the temperature at which switching occurs). Based on this fact, recent studies on $Mn_3Sn$ have discussed that another SOT switching mechanism exists in addition to the "intrinsic" mechanism explained by the application of the injected spin torque on the chiral antiferromagnetic order[24,28,33,49,51,63] (Fig. 1**c**). The "intrinsic" SOT switching occurs when $T_{\mathrm{switch}}$ is sufficiently lower than its transition (Néel) temperature $T_{\mathrm{trans}}$ (≈420 K for $Mn_3Sn$[13,56–58]). In contrast, a modified SOT switching scheme is triggered when the Joule heating raises the temperature of the $Mn_3Sn$ layer close to $T_{\mathrm{trans}}$, causing the antiferromagnetic order to destabilize[29,30] (Fig.1**c**). In this "temperature-assisted" SOT switching mechanism, the final magnetic state is determined by the spin torque injected near $T_{\mathrm{trans}}$ during the cooling of the magnetic layer. Switching cannot occur through this mechanism if the current generating the spin torque is stopped before the magnetic layer has cooled enough to restabilize the magnetic ordering. Therefore, the minimum time required for switching via the temperature-assisted mechanism is determined by the characteristic cooling time of the magnetic layer. For reported microfabricated $Mn_3Sn$-based structures, this cooling time is typically in the range of sub-microseconds[29,30], which is much longer than the picosecond timescales expected from antiferromagnetic dynamics[11,28,51]. We thus see that keeping the temperature of the antiferromagnetic layer below its transition temperature ($T_{\mathrm{switch}} < T_{\mathrm{trans}}$) and realizing switching "intrinsically", where the SOT is applied directly to the magnetic order, is essential for achieving picosecond-order processing.

To avoid confusion in the following discussion, here we define our usage of the terminology "intrinsic" and "temperature-assisted" in the context of SOT-induced switching. We will use the term "intrinsic" when referring to the case when magnetic order is present throughout the duration of the electrical write pulse (Fig. 1**c**, top diagram) and "temperature-assisted" when the Joule heating associated with the write pulse is sufficient to raise the device temperature above the magnetic transition temperature ($T_{\mathrm{switch}} < T_{\mathrm{trans}}$) and the final magnetic state is determined by the cooling process after removal of the pulse (Fig. 1**c**, bottom diagram). We emphasize that Joule heating and a finite temperature rise exist even when the $Mn_3Sn$ is switched by the intrinsic mechanism-indeed, such Joule heating will always be present when current is applied to an electrical device, however small it may be. The key distinction is that the elevated temperatures present during intrinsic switching are not enough to destroy the antiferromagnetic ordering in the intrinsic SOT switching process, whereas destabilizing the magnetic order using Joule heating is a requirement for switching by the temperature-assisted mechanism.

To achieve SOT switching via the intrinsic mechanism, it is crucial to minimize the temperature rise in the magnetic layer caused by the switching current pulse for both ferromagnets and antiferromagnets. The key lies in the thickness dependence of the switching current density. When the magnetic layer is accurately described by a macrospin model that absorbs all of the injected spin angular momentum (a good model for magnetic films thicker than the spin relaxation length $t_{\mathrm{rel}}$), it is known from studies on perpendicularly magnetized ferromagnetic films that the switching current density for the intrinsic mechanism ($j_{\mathrm{C}}^{\mathrm{int}}$) grows linearly with the thickness of the magnetic layer $t$[65,66]. This macrospin description breaks down when $t$ exceeds the exchange length $t_{\mathrm{ex}}$, at which point $j_{\mathrm{C}}^{\mathrm{int}}$ should saturate (Fig. 1**a**). (It should also be noted that for thick devices, as can be achieved by antiferromagnets, a similar saturation might be caused by the presence of grain boundaries, which may suppress angular momentum transfer.) This behavior of $j_{\mathrm{c}}^{\mathrm{int}}(t)$ can be understood through the conservation of angular momentum: the spin current required for switching must be proportional to the thickness/volume of the magnetic medium being switched. On the other hand, if switching is achieved by raising the temperature of the magnetic layer to $T_{\mathrm{trans}}$, and triggering the temperature-assisted mechanism, the required current density ($j_{\mathrm{C}}^{\mathrm{temp}}$) should scale with $t^{-0.5}$, independent of any other length scales (Supplementary Material SM9). In actual devices, the mechanism with the smaller switching current density should be selected, and a crossover of switching mechanisms driven by the thickness of the magnetic layer is expected at a crossover thickness $t_c$, where $j_{\mathrm{C}}^{\mathrm{temp}}(t_c) = j_{\mathrm{C}}^{\mathrm{int}}(t_c)$. In ferromagnetic systems, there is little room for changing the film thickness and thus for discussing SOT switching as a function of film thickness. On the other hand, in antiferromagnetic systems that offer a broader range of film thickness, a deeper understanding of SOT switching mechanisms can be constructed through experiments conducted with various film thicknesses to fully realize their potential.

Here we present experimental evidence of the crossover from the intrinsic mechanism to the temperature-assisted mechanism for the SOT switching of an antiferromagnetic order, employing polycrystalline films of the chiral antiferromagnet $Mn_3Sn$ as an example. By reducing the thickness of the $Mn_3Sn$ layer, we demonstrate that enhancing the expulsion of Joule heating from the device enables access to switching via the intrinsic SOT mechanism. Additionally, the thickness dependence further reveals that the exchange length in the antiferromagnet layer is at least on the order of tens of nanometers. These findings indicate that further miniaturization of antiferromagnetic devices could enable the predicted picosecond dynamics[27,28,51,63] in ultrafast memory applications.

## Results and Discussion

To investigate the thickness dependence of SOT switching in antiferromagnets, we employ $Mn_3Sn$($t$)/Ta(5 nm)/$AlO_x$(3 nm) films grown by magnetron sputtering on Si/$SiO_2$ (500 nm) substrates (Fig. 2**a**), where 15 nm$\leq t \leq$ 200 nm. Here, Ta acts as a source of spin current injected into the $Mn_3Sn$ layer through the SHE, while the $AlO_x$ layer protects the device from oxidation. It is known that composition can affect the magnetic properties of $Mn_3Sn$[67–69]. We prevent any systematic effects of composition in our series of films by growing films with varying thickness at once during the same deposition cycle where the thickness is controlled with a tilting shutter. The compositions of the films were determined to be $Mn_{3.06(2)}Sn_{0.94(2)}$ by scanning electron microscopy-electron dispersive X-ray spectroscopy independent of thickness. The films were annealed at 500°C after room temperature deposition of all layers.

We obtain high-quality film surfaces and $Mn_3Sn$/Ta interfaces with root mean square (RMS) values of 0.5~0.6 nm (obtained from atomic force microscopy (AFM) and transmission electron microscopy (TEM) images (Figs. 2**b** & 2**c**). This conclusion is also supported by X-ray reflectivity measurements, which complement the limitations of AFM and TEM in directly probing the roughness of the $Mn_3Sn$/Ta interface (Supplementary Material SM2)). This RMS value is significantly smaller than previous reports in similar $Mn_3Sn$ films[15,30], where the $Mn_3Sn$ layer was annealed before the deposition of subsequent layers. The improved interface, in turn, enables access to $Mn_3Sn$ films across a wide range of thicknesses down to 15 nm. We confirm through EDX mapping of the interfacial TEM image that the elemental diffusion between the $Mn_3Sn$/Ta is negligible, contained within a nanometer of the interface (Supplementary Material SM2).

X-ray diffraction (XRD) $2\theta/\omega$ scans show that while peaks corresponding to various crystal orientations are present in the spectra, the (0002) peak is significantly suppressed compared to what is expected from a completely randomly oriented sample (Figs. 2**d** & **e**). (The peaks from the $2\theta/\omega$ scan on the $t$=15 nm film could not be reliably calculated due to the weak signal coming from the small sample volume-see Supplementary Material SM4 for $2\theta\chi/\phi$ scans that supplement the weak signal from the thinner samples seen in the $2\theta/\omega$ scans). We thus conclude that, while retaining a degree of randomness, there exists a preference for the kagome planes to point out-of-plane for the entire range of thicknesses. This is in contrast to the randomly oriented polycrystalline $Mn_3Sn$ films deposited on similar substrates that were annealed without any capping[15,47,70,71]. Importantly, this preference for out-of-film-plane Kagome planes can be seen for all our samples regardless of thickness.

Hall measurements with a magnetic field $H$ in the out-of-plane direction reveal that our films have an enhanced anomalous Hall conductivity $\sigma_{yx}$ and smaller coercive field compared to identically stacked reference films where the $Mn_3Sn$ layer was annealed before the deposition of Ta and $AlO_x$ (Fig. 2**f**). The above characterizations of our samples suggest that the presence of immiscible capping layers during annealing can improve the quality of $Mn_3Sn$ thin films.

Our films were patterned into Hall bars with channel widths and lengths of 16 μm and 96 μm to perform electrical switching measurements (Fig. 3**a**). Fig. 3**b** shows the Hall resistance $R_H$ as a function of the magnitude of the "write" current pulses $I_{write}$ with a pulse duration of $\tau_{pulse}$ = 100 msat room temperature and a bias magnetic field $\mu_0 H_{bias}$ = 0.1 T applied along the current direction for the Hall bars fabricated from a $Mn_3Sn$(40 nm)/Ta(5 nm)/$AlO_x$(3 nm) film (Fig. 3**b**) (Methods). Henceforth, we call this stack structure "Config. 1". While we recognize that the presence of a bias field is impractical to implement on commercialized chips, we employ one in our experiments as an easy method to break the lateral symmetry breaking, which is required to observe deterministic SOT switching.

A switching of the sign of $R_H$ is observed when $I_{write}$ is large enough. This behavior qualitatively reproduces previous reports of current-induced switching in $Mn_3Sn$ films[24–26,28–30,32]. From these loops, we can define the switching current $I_C$ as the value of $I_{write}$ at which the transverse resistance changes sign[72–74]. $I_C$ is used to calculate the current density in the Ta layer ($j_C^{Ta}$). Note that this value can be defined without any knowledge of the mechanism causing the switching.

To show that the switching is induced by SOT, we performed the same switching experiment using a Ta(5 nm)/$Mn_3Sn$(40 nm)/$AlO_x$(3 nm) device (Fig. 3**b**, Config. 2). The polarity of the SOT should be reversed when the Ta-$Mn_3Sn$ stacking order is flipped, resulting in an inversion of the switching loop. We indeed confirm that the switching polarity in Config. 2 is opposite to that in Config. 1 while the electrically controllable Hall signal $\Delta R_H^{elec}$ (defined as $R_H(I_{write} = +0) - R_H(I_{write} = -0)$, where $I_{write} = +0$ indicates $I_{write} = 0$ mA while sweeping from positive to negative and vice versa) is almost identical (Fig. 3**b**). The polarity reversal caused by reversing the stacking order indicates that the final state of the current-induced magnetic switching is determined by spin–orbit torque (SOT), rather than by alternative mechanisms such as the Oersted field or self-generated spin-polarized currents that do not originate from spin currents generated by the heavy metal layer[43]. All following data on switching experiments are obtained using the film structure in Config. 1.

As discussed above, the thickness dependence of the switching current density may reveal the mechanism behind the SOT switching in $Mn_3Sn$. To verify this in our samples, we plot $j_C^{Ta}$ against $t$ in Fig. 4**a**. We observe a decreasing behavior in $j_C^{Ta}$ with increasing $t$ in the thicker film regime ($t > 30$ nm). For the temperature-assisted switching mechanism, we anticipate a power law scaling of $j_C^{Ta} \propto t^{-0.5}$ (Supplementary Material SM9). The pink curve in Fig. 4**a** represents a fit to $j_C^{Ta} = at^{-0.5}$ ($a = \text{const.}$) for $t > 60$ nm, which we see well describes the data for thicker devices, while a significant deviation in preference for a smaller $j_C^{Ta}$ is seen below the crossover thickness $t_c \sim 30$ nm. To demonstrate that this deviation is not due to the current shunting in the Ta layer, we plot the critical switching power $P_C = I_C V_C$ as a function of $t$ in Fig. 4**b**, where $V_C$ is the voltage applied to the device to generate the switching current $I_C$. $P_C$ has been reported to be independent of $t$ in the temperature-assisted switching[29], which we also confirm through numerical simulations even in the presence of current shunting into the Ta layer (Supplementary Material SM9). In contrast, a clear decreasing trend in $P_C$ is found below $t_c$, the thickness at which $j_C^{Ta}$ deviates from the power scaling $\propto t^{-0.5}$. We can, therefore, interpret these results as signatures of lower switching temperatures in thinner devices ($t < t_c$), indicating the presence of intrinsic SOT switching. The small $P_C$ in the thinner devices suggests the superior energy efficiency of the intrinsic SOT switching.

We emphasize that the value of $t_c$ observed in our set of devices is not determined simply by the parameters intrinsic to $Mn_3Sn$. Rather, it is determined by the competition between the amount of Joule heating versus the rate at which heat can escape from the device. Many parameters can affect this balance, such as the physical properties of the film, including electrical resistivity and heat capacity, the substrate material, the thermal contact between the device and the substrate, and the device dimensions. On the other hand, as we discuss below, $t_{ex}$ is at least 30 nm, enabling observations of the crossover between the intrinsic SOT regime and the temperature-assisted one across $t_c$.

The size of the $Mn_3Sn$ layer that can be manipulated by SOT can be evaluated through the switching ratio $\xi = \left| \Delta R_H^{elec} / \Delta R_H^{field} \right|$, where $\Delta R_H^{field}$ is defined as $R_H(H = -0) - R_H(H = +0)$ under an out-of-plane magnetic field ($H$) sweep ($H = +0$ indicates $\mu_0 H = 0$ T while sweeping from positive to negative and vice versa). The obtained values of $\xi$ as a function of $t$ are shown in Fig. 4**c**. In the regime $t \leq t_c \sim 30$ nm, where intrinsic switching is dominant, $\xi$ shows little dependence on $t$ and is consistently sizeable (~60 %), while it rapidly decreases in our thicker films associated with temperature-assisted switching. These results experimentally indicate that our $Mn_3Sn$ films have an exchange

length of at least on the order of $t_c \sim 30$ nm and that antiferromagnetic devices with such thicknesses can indeed be switched intrinsically. The suppressed $\xi$ in the $t = 15$ nm sample could be due to the relatively poor crystallinity of the $Mn_3Sn$ layer of this particular film. (See Supplementary Material SM11 for further discussion of the thickness dependence of $\xi$.)

To further investigate the effect of film thickness on the device temperature during switching, we performed electrical switching measurements under multiple sample stage temperatures $T_{\mathrm{stage}}$. As Joule heating is known to follow the current quadratically ($\propto I^2$), we plot $I_{\mathrm{C}}^2$ against $T_{\mathrm{stage}}$ for devices with various $t$ in Fig. 5**a**. For thicker devices ($t \geq t_c \sim 30$ nm), we see that $I_{\mathrm{C}}^2$ and $T_{\mathrm{stage}}$ lie well on top of a straight line in the temperature range $T_{\mathrm{stage}} \leq 380$ K. This indicates that the temperature rise in the device, $T^* - T_{\mathrm{stage}}$, should be proportional to $I_{\mathrm{C}}^2$. Here, $T^*$ is a characteristic temperature of the device during the application of the switching current $I_{\mathrm{C}}$. $T^*$ can be extracted by fitting $I_{\mathrm{C}}^2$ with the relation $I_{\mathrm{C}}^2 = \lambda(T^* - T_{\mathrm{stage}})$ where $\lambda$ and $T^*$ are the fitting parameters ($\lambda$ contains information about the heat capacity of the device). We see that the fits cross the $T_{\mathrm{stage}}$-axis near 420 K, corresponding to $T^* \sim 420$ K. This is consistent with the temperature-assisted switching mechanism, where deterministic switching occurs when the device temperature reaches $T_{\mathrm{trans}} \sim 420$ K[29,30]. In contrast, when $t \leq t_c \sim 30$ nm, the values of $I_{\mathrm{C}}^2$ can be fit by the above model with a value of $T^*$ less than 420 K, i.e., $T_{\mathrm{trans}}$, in the entire range of $T_{\mathrm{stage}}$. These results suggest that switching in the devices with $t \leq t_c \sim 30$ nm is achieved without heating the device up to $T_{\mathrm{trans}}$ even when $T_{\mathrm{stage}}$ is at room temperature. Moreover, it should be noted that the values of $I_{\mathrm{C}}^2$ measured at $T_{\mathrm{stage}} \geq 380$ K fall below what is expected from the linear fit obtained from data points at lower temperatures. This result indicates that switching occurs below $T_{\mathrm{trans}}$ at $T_{\mathrm{stage}} > 380$ K, similar to the thinner films. This can be attributed to the increase in base temperature reducing the magnetic anisotropy, effectively lowering $j_{\mathrm{C}}^{\mathrm{int}}$ below $j_{\mathrm{C}}^{\mathrm{temp}}$.

By using the linear model described above to fit the measured values of $I_{\mathrm{C}}^2$ in the range $T_{\mathrm{stage}} \leq 380$ K, we see that $T^*$, visualized as the horizontal intercepts of the fitted lines in Fig. 5**a**, shifts below 420 K with decreasing $t$ below $t_c \sim 30$ nm. The values of $T^*$ for various $t$ are presented in Fig. 5**b**, along with the experimental values of $T_{\mathrm{trans}}$ obtained by measuring the temperature evolution of AHE (Supplementary Material SM12). Fig. 5**b** confirms that $T^*$ moves to smaller temperatures and away from $T_{\mathrm{trans}}$ in thinner devices, while at larger $t$, $T_{\mathrm{trans}}$ and $T^*$ saturate at similar values. We note that the thickness range in which we find $T^* < T_{\mathrm{trans}}$, $t \leq t_c$, overlaps with the thickness range in which the room temperature values of $J_C^{Ta}$ and $P_C$ deviate from the behavior predicted for temperature-assisted switching (Figs. 4**a** & 4**b**). These observations further support our claim that we are changing the dominant mechanism for the switching behavior as a function of the film thickness across $t_c \sim 30$ nm. For completeness, we show in Fig. 5**c** the calculated steady-state temperature $T_{\mathrm{steady}}$ for devices of various $t$ upon application of the experimentally observed values for $I_C$. The calculations qualitatively reproduce the crossover behavior of the switching temperature $T^*$ under the critical current across $t_c$ ~30 nm. All our experimental observations and analyses reveal that the intrinsic mechanism dominates for the films with thickness $t \leq t_c$~30 nm.

In the temperature-assisted switching mechanism, the switching time has been reported to be limited to timescales on the order of sub-μs and thus attempts to achieve switching with pulses shorter than several hundred nanoseconds have failed[29,30]. To clearly confirm the intrinsic SOT switching in the thinner film regime, we performed switching experiments on devices with $t$ = 15 nm using write pulse currents with durations $\tau_{\mathrm{pulse}} \leq$ 50 ns with nanosecond-order rise/fall times, much shorter than the estimated value (~200 ns) of the cooling times of the device (Supplementary Material SM10). The switching measurements were performed at room temperature with $\mu_0 H_{\mathrm{bias}}$ = 0.1 T. As shown in Fig. 6**a**, we observe switching in our $t$ = 15 nm devices with a $\tau_{\mathrm{pulse}}$ as short as 10 ns. Importantly, the switching ratio $\xi$ is essentially unchanged from those obtained with 100 ms pulses with rise and fall times of <1 ms, demonstrating that switching with such short pulse currents should prevent the temperature-assisted switching and ensure that the antiferromagnetic phase is preserved for the entire duration of the current pulse. Our results also further indicate that the exchange length should be at least on the same order of the film thickness, i.e., 15 nm.

Finally, we demonstrate that switching in $Mn_3Sn$ in our $t$=15 nm samples can be achieved by single-shot pulses as short as 10 ns. We first initialize the magnetic state using a 100 ms-pulse with amplitude above the switching threshold as determined in the switching loops, after which we inject a 10 ns-pulse with amplitude of 13.5 V, sufficient to achieve switching as determined by the data in Fig. 6**a**. The Hall resistance of the device after the injection of both the 100 ms reset pulse and the 10 ns pulse, measured with a weak DC current of 0.2 mA, is shown in Fig, 6**b**. We see that a single 10 ns pulse is sufficient to achieve a change in the Hall resistance of ~1 Ω, equivalent to the full switching amplitude seen in the data in Fig. 6**a**. The observation of switching of the antiferromagnetic order by a single nanosecond-order pulse demonstrates the potential for ultrafast operation of $Mn_3Sn$-based memory by exploiting the intrinsic switching mechanism.

In conclusion, we demonstrate that the mechanism behind the electrical switching transitions from the temperature-assisted mechanism to the intrinsic one as the antiferromagnetic $Mn_3Sn$ layer becomes thinner. This crossover occurs when the switching current density required for intrinsic switching becomes smaller than that required to heat the device to $T_{trans}$. Furthermore, the experimentally obtained exchange length of at least 30 nm, indicated by the crossover thickness, shows that in functional antiferromagnets such as $Mn_3Sn$, proper management of Joule heating should allow magnetic switching via the intrinsic mechanism in antiferromagnetic devices with much thicker magnetic layers than those in perpendicularly magnetized ferromagnetic devices. Employing the intrinsic mechanism for SOT-driven switching represents a crucial step towards realizing ultrafast information processing in the picosecond (ps) range in antiferromagnets.

In this study, we used the thickness of the antiferromagnetic layer as our control parameter; however, we anticipate that further optimization of the heat transfer from the device to the substrate through other variables, such as the substrate material and device dimensions, will enable antiferromagnetic SOT switching at much lower temperatures. As we advance towards memory applications for antiferromagnetic materials, further thinning of the magnetic layers and miniaturization of the devices will continue, enhancing the functionalities of the intrinsic SOT mechanism. This progress is expected to lead to even faster and more energy-efficient information processing.

## Methods

### Sample preparation

$Mn_3Sn$/Ta/$AlO_x$ multilayers were deposited by magnetron sputtering onto a Si substrate with a thermally oxidized $SiO_2$ layer (500 nm). The target-to-substrate (T/S) distance was 20 cm for $Mn_3Sn$ and Ta, and 15 cm for $AlO_x$. The Ta and $AlO_x$ thicknesses were fixed at 5 and 3 nm, respectively, while the $Mn_3Sn$ thickness $t$ was varied from 15 to 200 nm. All layers were deposited at room temperature with a base pressure of $<1\times10^{-6}$ Pa, and the stacks were subsequently annealed for 30 minutes at 500°C. Other growth processes were used for reference purposes and are detailed in the text where relevant. Standard photolithography and Ar ion-milling techniques were used to pattern the Hall bars. Ti (5 nm)/Au (200 nm) contact pads for the Hall bars were deposited by electron beam evaporation and patterned using conventional liftoff processes.

## Sample characterization

The composition of the $Mn_3Sn$ layers was determined as $Mn_{3.06(2)}Sn_{0.94(2)}$ by scanning electron microscopy-energy dispersive X-ray spectrometry. The thickness of each layer was calculated using pre-calibrated deposition rates obtained by x-ray reflectivity analysis performed with a Rigaku Smartlab unit with a Cu-K$\alpha$ source (wavelength: 1.541 Å). X-ray diffraction (XRD) measurements were conducted using the same Rigaku Smartlab unit. All $2\theta/\omega$ XRD spectra was normalized based on the (400) Si substrate peak appearing near $2\theta \approx 69.13°$ used to align the sample with respect to the optical components during measurement.

Simulated XRD spectra were calculated using VESTA[77] based on a cif file of $Mn_3Sn$, which specifies the space group of the crystal structure ($P6_3$/mmc) and the lattice parameters ($a$=$b$=5.67 Å, $c$=4.53 Å). All relevant reflections and structure factors were computed using the crystal structure and lattice parameters. Peak intensities were calculated assuming a randomly oriented powder sample and a Cu-Kα X-ray wavelength ($\lambda$ = 1.541 Å).

Atomic force microscopy (AFM) and cross-sectional transmission electron microscopy (TEM) images were collected at room temperature.

Details on the characterization of the magneto-transport properties of our samples can be found in Supplementary Material SM1.

## Electrical switching measurements

To achieve electrical switching, a DC current pulse ("write pulse") $I_{\mathrm{write}}$ with duration $\tau_{\mathrm{pulse}}$ was injected into a Hall bar. $\tau_{\mathrm{pulse}}$ was set to be 100 ms with rise and fall times of <1 ms except in the sub-100 ns switching measurements presented in Fig. 6, in which the rise and fall times were a few nanoseconds. The transverse voltage $V_{yx}$ was subsequently measured after a delay time of 600 ms using a small DC current $I_{\mathrm{read}}$ =0.2 mA. Any subsequent write pulses were applied only after waiting for 1 s from the time that the previous read pulse was turned off. A bias magnetic field $\mu_0 H_{\mathrm{bias}}$ of 0.1 T was applied parallel to the current direction to break lateral symmetry. All room-temperature switching measurements were performed in air, while the stage temperature dependence presented in Fig. 5 was measured in a He environment. All magnetic field- and $I_{\mathrm{write}}$- even signals in the anomalous Hall and electrical switching loops are subtracted in the presented data.

The switching current $I_{\mathrm{C}}$ is defined as the value $I_{\mathrm{write}}$ at which the sign of $R_{\mathrm{H}}$ changes in the switching loops. $j_{\mathrm{C}}^{\mathrm{Ta}}$ is the current density in the Ta layer under the application of current $I_{\mathrm{C}}$ and was calculated using a two-resistor model using $\rho_{\mathrm{Ta}}$ = 200 μΩ cm[78] as the resistivity of the Ta layer and the measured longitudinal resistivity of the Hall bar.

### Numerical simulations

The details of the numerical simulations used in this study are outlined in Supplementary Material SM7.

## Author contributions

T. Matsuo and T.H. contributed equally to this work. S.N. and T.H. planned the project. T. Matsuo and T.H. fabricated the thin films. T. Matsuo, D.N.-H., and R.U. characterized the thin films. T. Matsuo and T.H. fabricated the devices with help from H.T., K.K. and Y.O. | T. Matsuo, T.H., and T. Matsuda performed the transport and switching measurements with help from H.T. and S.M. | T. Matsuo performed numerical simulations with help from H.T. | T. Matsuo, T.H., and S.N. wrote the manuscript with input from S.M. All authors discussed the results and commented on the manuscript.

## Competing interests

The authors declare no competing interests.

## Acknowledgements

We thank S. Yamada for discussions. This work was partially supported by JST-MIRAI Program (No. PMJMI20A1), JST-ASPIRE(JPMJAP2317), and MEXT/JSPS-KAKENHI (Nos. 15H05882, 15H05883, 15K21732, 19H00650, JP25H01250). A part of this work was conducted at Takeda Sentanchi Supercleanroom, The University of Tokyo, supported by the Nanotechnology Platform Program of the Ministry of Education, Culture, Sports, Science and Technology (MEXT), Japan (Nos. JPMXP09F20UT0141, JPPMXP09F21UT0141, and JPMXP09F22UT1181). The Institute for Quantum Matter, an Energy Frontier Research Center, was funded by DOE, Office of Science, Basic Energy Sciences under Award No. DE-SC0024469. T.H. acknowledges support from JST-PRESTO (No. JPMJPR24M8) and the Murata Science Foundation. The use of the facilities of the Laboratory for Magnetic and Electronic Properties at Interface and the Materials Design and Characterization Laboratory at the Institute for Solid State Physics, the University of Tokyo, are acknowledged.

## Figure captions

**Fig. 1| SOT-switching modes of $Mn_3Sn$. a**, (Top) Schematics of the size of the magnetic order parameter during the switching process for thin (left) and thick (right) films. (Bottom) Schematics of the expected behavior in the switching current density as a function of the device thickness. In intrinsic SOT switching, the switching current density $j_{\mathrm{C}}^{\mathrm{int}}$ should increase linearly with the magnetic layer thickness $t$ above the spin relaxation length $t_{\mathrm{rel}}$until saturating around the exchange length $t_{\mathrm{ex}}$ (blue). The switching current density for temperature-assisted SOT switching $j_{\mathrm{C}}^{\mathrm{temp}}$ is expected to scale with $t^{-0.5}$ (red). The magnitude of the $j_{\mathrm{C}}^{\mathrm{temp}}$ curve is determined by the rate of heat loss from the device to the substrate and the value of the magnetic transition temperature $T_{\mathrm{trans}}$. **b**, (Left) Crystal and magnetic structures of $Mn_3Sn$. The light and dark blue spheres represent Mn atoms, while the black and white spheres represent Sn atoms at the atomic positions $z$ =0 and 1/2, respectively. Figure generated by VESTA[77]. (Right) View of the cluster comprised of six Mn moments used to define the cluster magnetic octupole (purple). **c**, Schematics of intrinsic and temperature-assisted SOT switching in $Mn_3Sn$. In the intrinsic mechanism, the chiral antiferromagnetic order is intact during the application of current and switches once the current density reaches a threshold value. On the other hand, the antiferromagnetic order is significantly weakened during current application in the temperature-assisted picture. The final state is determined by the polarity of the injected spin torque as the antiferromagnetic layer cools below $T_{\mathrm{trans}}$ while the current is being removed.

**Fig. 2| Characterization of the $Mn_3Sn$/Ta films. a,** Schematic illustration of the $Mn_3Sn$/Ta/$AlO_x$ multilayers used in this study. **b**, Atomic force microscopy (AFM) image of a $Mn_3Sn$(40 nm)/Ta(5 nm)/$AlO_x$(3 nm) film. The surface RMS roughness was calculated to be 0.6 nm. **c**, Cross-sectional transmission electron microscopy (TEM) image of a $Mn_3Sn$(40 nm)/Ta(5 nm)/$AlO_x$(3 nm) film. The layer boundaries (white lines) were generated by the Canny edge detection algorithm[79] and yielded a RMS roughness value of ~ 0.5 nm for the $Mn_3Sn$/Ta interface. **d**, $2\theta/\omega$ X-ray diffraction (XRD) signals

measured for $Mn_3Sn(t)/Ta(5\ nm)/AlO_x(3\ nm)$ films on Si(380 μm)/$SiO_2$(500 nm) substrates with 15 nm $\leq t \leq$ 200 nm. The simulated data at the bottom assumes a completely random crystal orientation. The peaks in the region marked with the asterisk are from the substrate. **e,** Comparison of the relative intensities of the $(02\bar{2}0)$, (0002), and $(02\bar{2}1)$ XRD peaks measured by the $2\theta/\omega$ geometry. **f**, Anomalous Hall conductivity $\sigma_{yx}$ as a function of perpendicular magnetic field $H$ for two $Mn_3Sn$(40 nm)/Ta(5 nm)/$AlO_x$(3 nm) films. The results for the film that was annealed at 500°C after room temperature deposition of all layers are shown in blue ("Annealed with capping"), while the pink curve ("Reference") shows the data for the film where the $Mn_3Sn$ layer was annealed at 500°C before the deposition of the Ta and $AlO_x$ layers.

**Fig. 3| SOT switching in $Mn_3Sn$/Ta bilayers at room temperature. a**, Photograph of a Hall bar used in electrical switching measurements. The channel length and width are 16μm and 96 μm, respectively. **b**, (Left) Schematics of current switching in our films. Note that in Configs. 1 ($Mn_3Sn$(40 nm)/Ta(5 nm)/$AlO_x$(3 nm)) and 2 (Ta(5 nm)/ $Mn_3Sn$(40 nm)/$AlO_x$(3 nm)), the polarity of the spin current being injected in the $Mn_3Sn$ layer is opposite, which should result in a reversal of the switching loop. (Right) $R_{\mathrm{H}}$ as a function of $I_{\mathrm{write}}$ for a $Mn_3Sn$(40 nm)/Ta(5 nm)/$AlO_x$(3 nm) and Ta(5 nm)/$Mn_3Sn$(40 nm)/$AlO_x$(3 nm) film measured at room temperature. The bias field was fixed to be $\mu_0 H_{\mathrm{bias}} = 0.1$ T.

**Fig. 4| Crossover of SOT switching mechanism as a function of $Mn_3Sn$ thickness. a**, The switching current density in the Ta layer ($j_{\mathrm{C}}^{\mathrm{Ta}}$) plotted against the $Mn_3Sn$ layer thickness $t$. The pink line indicates the scaling behavior $j_{\mathrm{C}}^{\mathrm{Ta}} \propto t^{-0.5}$ in thicker devices with $t \geq t_c \sim 30$ nm. **b**, The critical switching power $P_{\mathrm{C}}$ as a function of $Mn_3Sn$ thickness $t$. $P_{\mathrm{C}}$ exhibits a decreasing behavior at $t \lesssim t_c \sim 30$ nm, indicating that the switching temperature in this regime is below $T_{\mathrm{trans}}$. The solid line represents a constant $P_{\mathrm{C}}$ with respect to $t$, which is expected if temperature-assisted switching is dominant in all thicknesses. (The value of $P_{\mathrm{C}}$ indicated by the solid line is the average value of the $t >$ 30 nm datapoints.) **c**, Switching ratio $\xi$ as a function of the $Mn_3Sn$ layer thickness $t$.

**Fig. 5| Switching measurements as a function of sample stage temperature. a,** $I_{\mathrm{C}}^2$ as a function of $T_{\mathrm{stage}}$ for devices fabricated from with $Mn_3Sn(t)$/Ta(5 nm)/$AlO_x$(3 nm) films with $t$ from 20 nm to 200 nm. The solid lines represent the linear fits obtained by the data points with $T_{\mathrm{stage}} \leq 380$ K. The values of $I_{\mathrm{C}}^2$ are normalized by their values at $T_{\mathrm{stage}} =$ 300 K to allow direct comparison between those for different $t$. The inset zooms into the

linear fits near 420 K. **b,** Values of $T^*$ (obtained from the fits $I_C^2 = \lambda(T^* - T_{\mathrm{stage}})$ shown in Fig. 5**a**)) and Néel temperature $T_{\mathrm{trans}}$ (obtained from temperature-dependent AHE measurements; see Supplementary Material SM12) in $Mn_3Sn(t)$/Ta(5 nm)/$AlO_x$ (3 nm) devices as a function of $t$. $T^*$ is relatively constant near $T_{\mathrm{trans}}$ across thicker devices with $t > t_c \sim 30$ nm while it is reduced with $t$ in thinner devices with $t \leq t_c \sim 30$ nm. In contrast, $T_{\mathrm{trans}}$ remains nearly the same within the entire range of $t$ used in our measurements. **c,** Calculated steady state temperatures for $Mn_3Sn(t)$/Ta(5 nm) devices on Si(380 μm)/$SiO_2$(500 nm) substrates under the observed critical current $I_C$ as a function of $t$. Heat transfer to the surrounding air was described by a heat transfer coefficient of 5 $W/m^2K$.

**Fig. 6| SOT switching of $Mn_3Sn$ with sub-100 ns pulses. a**, Observation of switching in a $Mn_3Sn$(15 nm)/Ta(5 nm)/$AlO_x$(3 nm) Hall deice at room temperature with $\mu_0 H_{\mathrm{bias}} = 0.1$ T using write pulses with $\tau_{\mathrm{pulse}}$ from 100 ms to 10 ns. The sub-100 ns electric pulses have rise and fall times of several nanoseconds. Inset: Switching ratio as a function of $\tau_{\mathrm{pulse}}$ for the $t$ = 15 nm device. **b,** Switching in $Mn_3Sn$(15 nm)/Ta(5 nm)/AlOx(3 nm) by 10 ns single-shot pulses with current densities $j^{\mathrm{Ta}}$ =40 $MA/cm^2$. The magnetic state is reset by a 100 ms pulse before the injection of the 10 ns pulse.

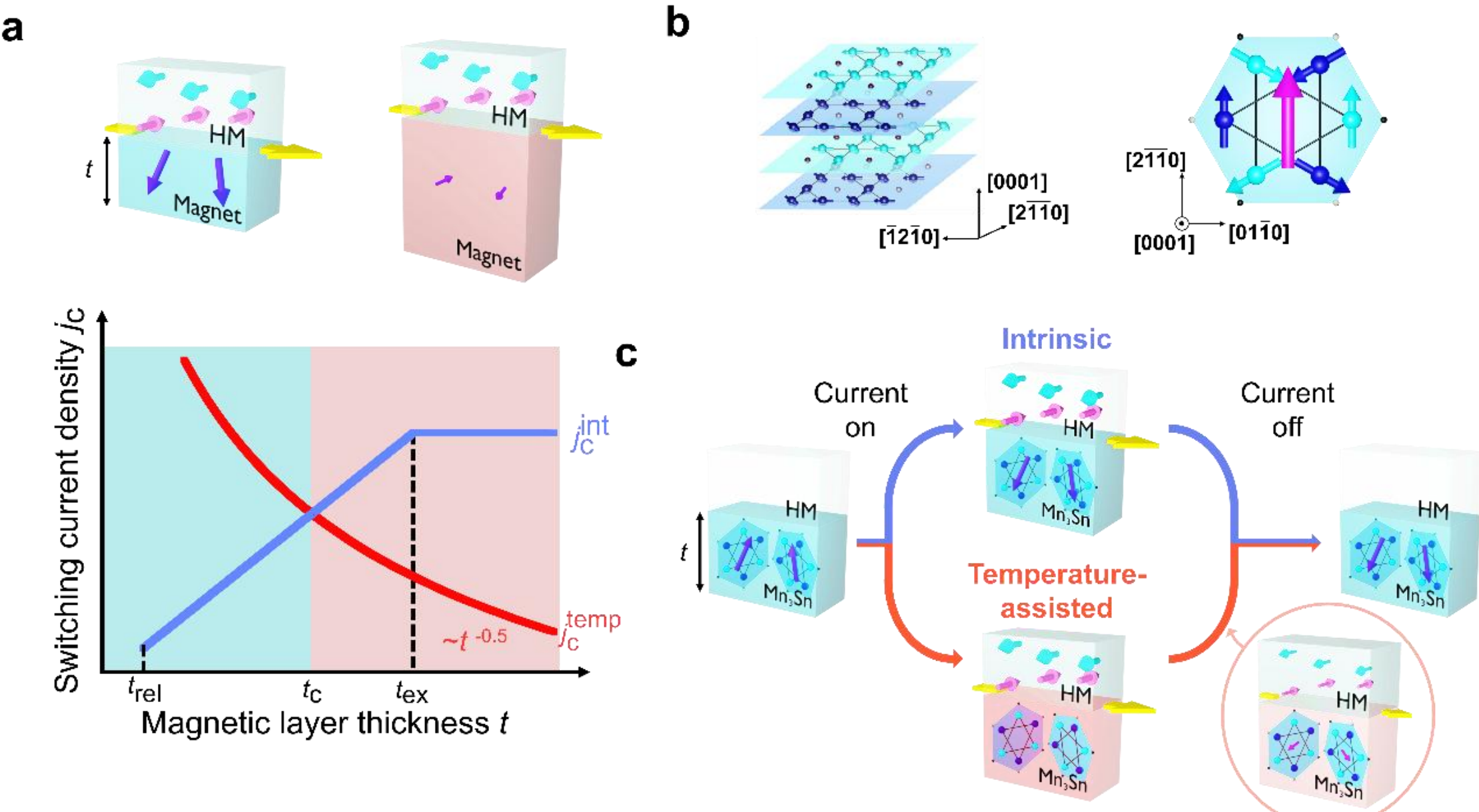

Figure 1

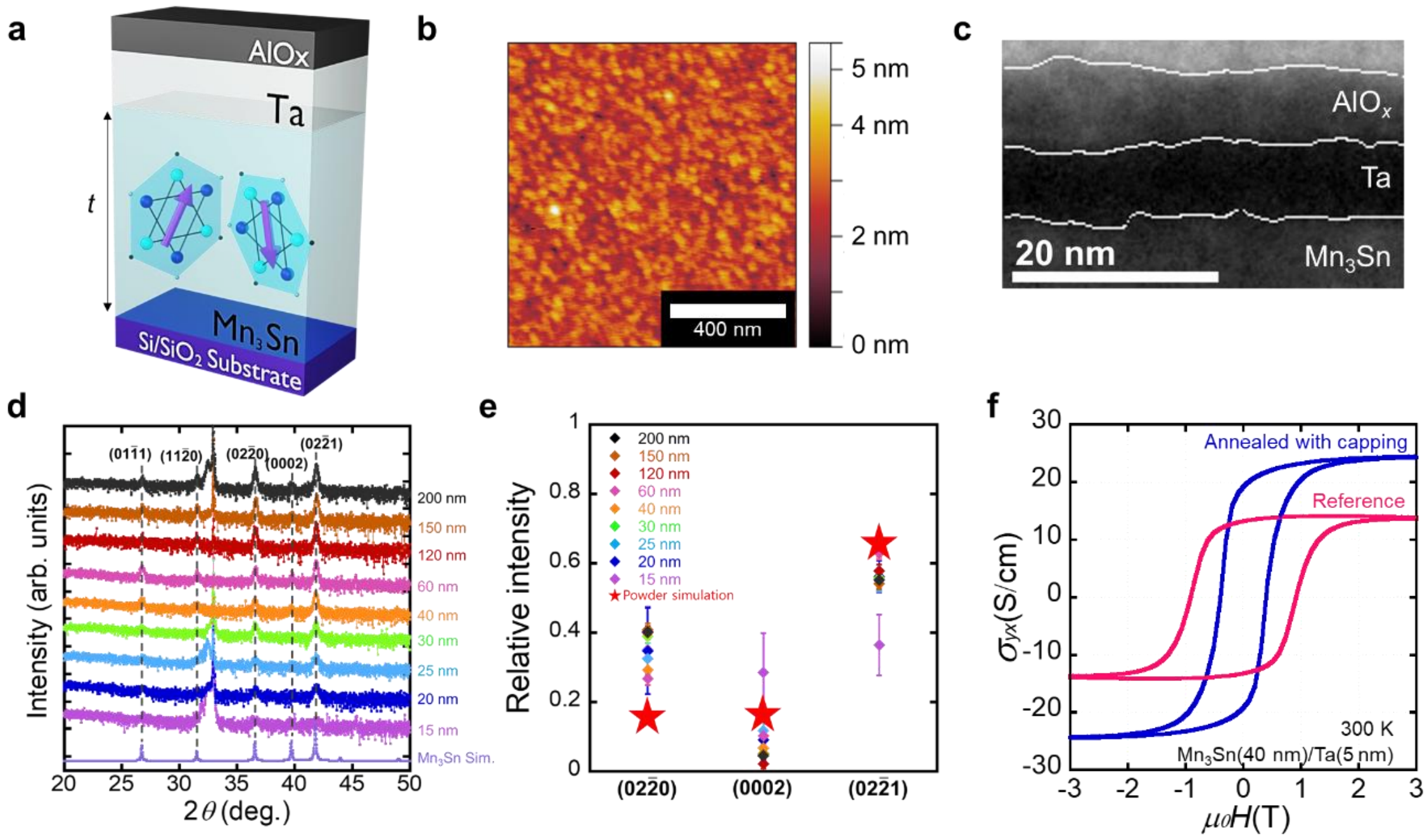

Figure 2

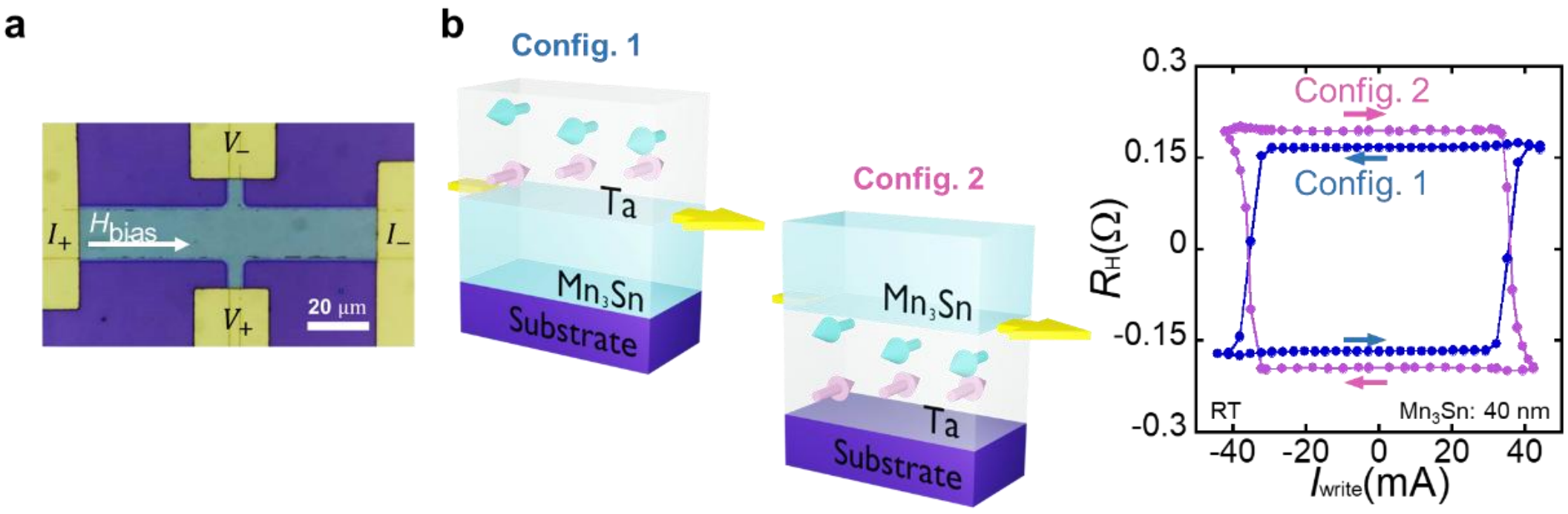

Figure 3

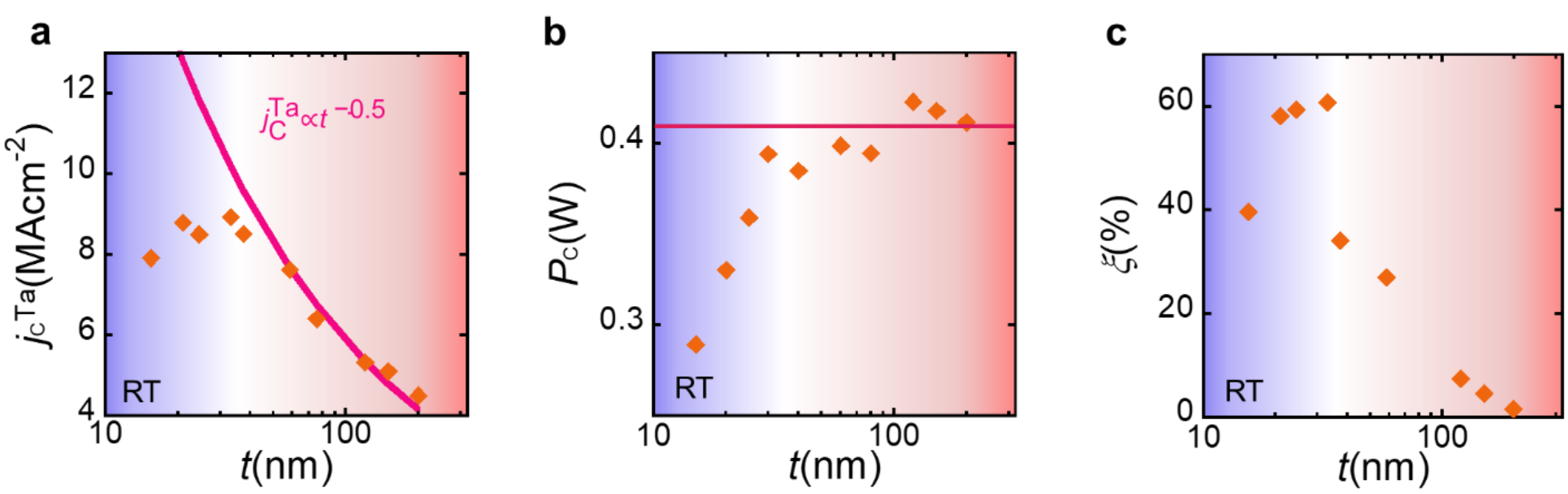

Figure 4

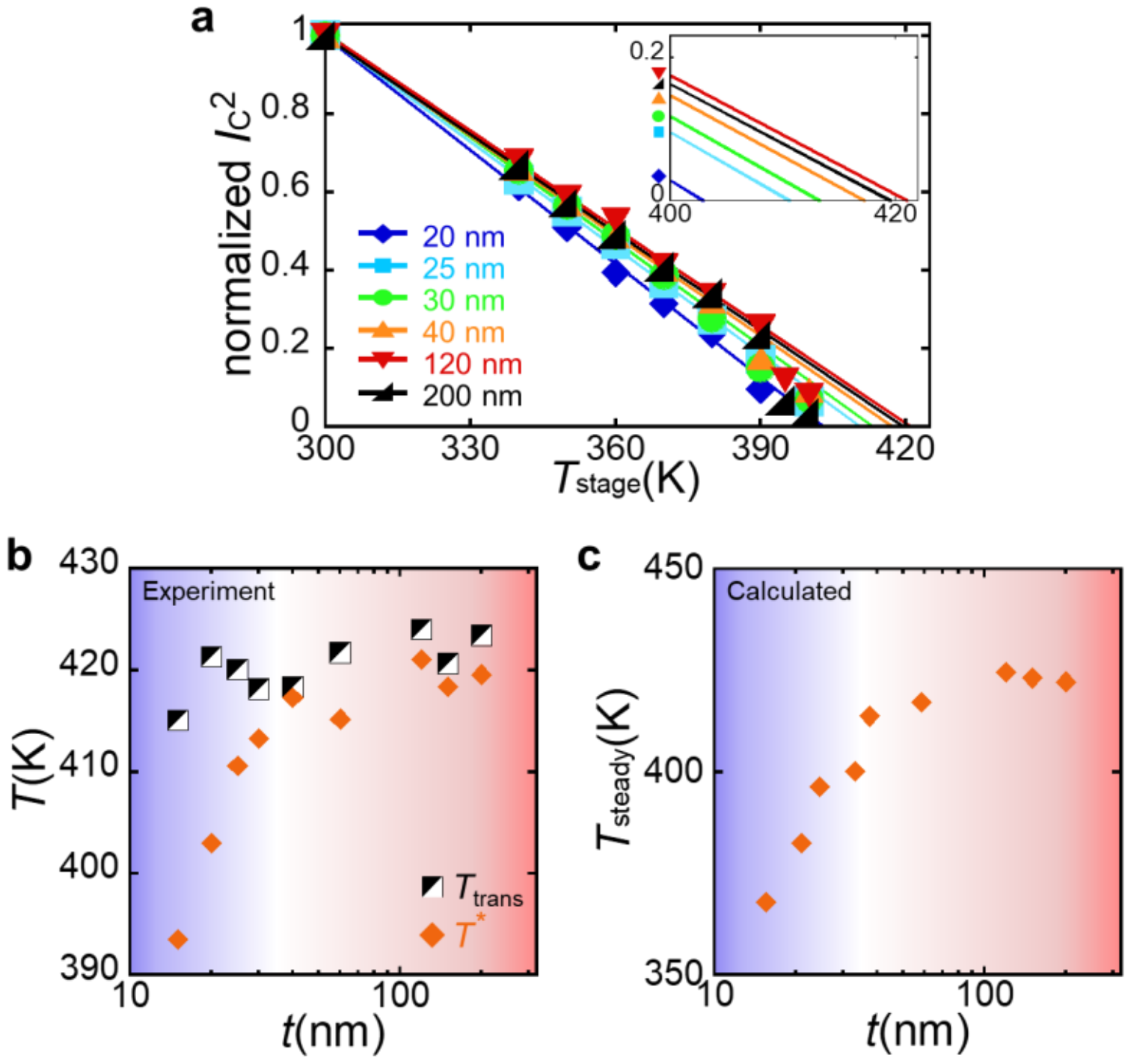

Figure 5

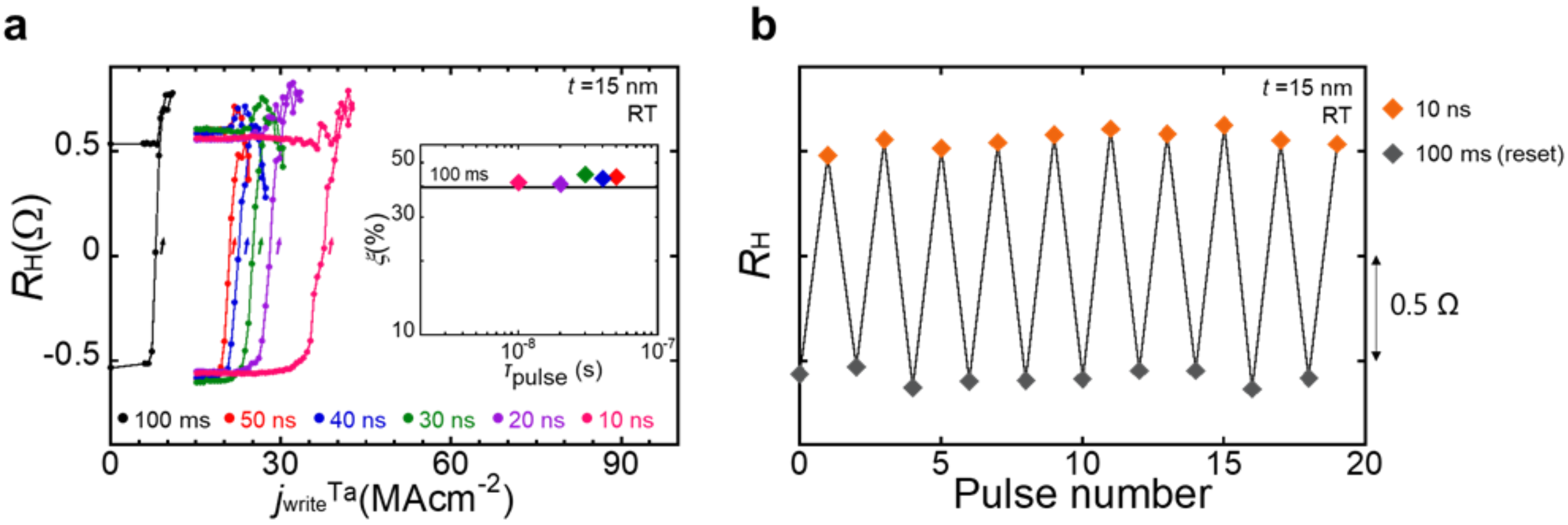

Figure 6

## References

1. Wolf, S. A., Lu, J., Stan, M. R., Chen, E. & Treger, D. M. The Promise of Nanomagnetics and Spintronics for Future Logic and Universal Memory. *Proc. IEEE* **98**, 2155–2168 (2010).
2. Bhatti, S. *et al.* Spintronics based random access memory: a review. *Mater. Today* **20**, 530–548 (2017).
3. Dieny, B. *et al.* Opportunities and challenges for spintronics in the microelectronics industry. *Nat. Electron.* **3**, 446–459 (2020).
4. Manchon, A. *et al.* Current-induced spin-orbit torques in ferromagnetic and antiferromagnetic systems. *Rev. Mod. Phys.* **91**, 035004 (2019).
5. Shao, Q. *et al.* Roadmap of Spin–Orbit Torques. *IEEE Trans. Magn.* **57**, 1–39 (2021).
6. Kittel, C. Theory of Antiferromagnetic Resonance. *Phys. Rev.* **82**, 565–565 (1951).
7. Keffer, F. & Kittel, C. Theory of Antiferromagnetic Resonance. *Phys. Rev.* **85**, 329–337 (1952).
8. Gomonay, E. V. & Loktev, V. M. Spintronics of antiferromagnetic systems (Review Article). *Low Temp. Phys.* **40**, 17–35 (2014).

9. Jungwirth, T., Marti, X., Wadley, P. & Wunderlich, J. Antiferromagnetic spintronics. *Nat. Nanotech.* **11**, 231–241 (2016).

10. Kirilyuk, A., Kimel, A. V. & Rasing, T. Ultrafast optical manipulation of magnetic order. *Rev. Mod. Phys.* **82**, 2731–2784 (2010).

11. Miwa, S. *et al.* Giant Effective Damping of Octupole Oscillation in an Antiferromagnetic Weyl Semimetal. *Small Sci.* **1**, 2000062 (2021).

12. Otani, Y. & Higo, T. Domain structure and domain wall dynamics in topological chiral antiferromagnets from the viewpoint of magnetic octupole. *Appl. Phys. Lett.* **118**, 040501 (2021).

13. Nakatsuji, S., Kiyohara, N. & Higo, T. Large anomalous Hall effect in a non-collinear antiferromagnet at room temperature. *Nature* **527**, 212–215 (2015).

14. Kuroda, K. *et al.* Evidence for magnetic Weyl fermions in a correlated metal. *Nat. Mater.* **16**, 1090–1095 (2017).

15. Higo, T. *et al.* Anomalous Hall effect in thin films of the Weyl antiferromagnet $Mn_3Sn$. *Appl. Phys. Lett.* **113**, 202402 (2018).

16. You, Y. *et al.* Room temperature anomalous Hall effect in antiferromagnetic $Mn_3SnN$ films. *Appl. Phys. Lett.* **117**, 222404 (2020).

17. Chen, T. *et al.* Anomalous transport due to Weyl fermions in the chiral antiferromagnets $Mn_3X$, $X$ = Sn, Ge. *Nat. Commun.* **12**, 572 (2021).

18. Šmejkal, L., MacDonald, A. H., Sinova, J., Nakatsuji, S. & Jungwirth, T. Anomalous Hall antiferromagnets. *Nat. Rev. Mater.* **7**, 482–496 (2022).

19. Feng, Z. *et al.* An anomalous Hall effect in altermagnetic ruthenium dioxide. *Nat. Electron.* **5**, 735–743 (2022).

20. Nakatsuji, S. & Arita, R. Topological Magnets: Functions Based on Berry Phase and Multipoles. *Annu. Rev. Condens. Matter Phys.* **13**, 119–142 (2022).

21. Gonzalez Betancourt, R. D. *et al.* Spontaneous Anomalous Hall Effect Arising from an Unconventional Compensated Magnetic Phase in a Semiconductor. *Phys. Rev. Lett.* **130**, 036702 (2023).

22. Reichlova, H. *et al.* Observation of a spontaneous anomalous Hall response in the $Mn_5Si_3$ d-wave altermagnet candidate. *Nat. Commun.* **15**, 4961 (2024).

23. Hajiri, T., Ishino, S., Matsuura, K. & Asano, H. Electrical current switching of the noncollinear antiferromagnet $Mn_3GaN$. *Appl. Phys. Lett.* **115**, 052403 (2019).

24. Tsai, H. *et al.* Electrical manipulation of a topological antiferromagnetic state. *Nature* **580**, 608–613 (2020).

25. Tsai, H. *et al.* Large Hall Signal due to Electrical Switching of an Antiferromagnetic Weyl Semimetal State. *Small Sci.* **1**, 2000025 (2021).

26. Tsai, H. *et al.* Spin–orbit torque switching of the antiferromagnetic state in polycrystalline $Mn_3Sn$/Cu/heavy metal heterostructures. *AIP Adv.* **11**, 045110 (2021).

27. Takeuchi, Y. *et al.* Chiral-spin rotation of non-collinear antiferromagnet by spin–orbit torque. *Nat. Mater.* **20**, 1364–1370 (2021).

28. Higo, T. *et al.* Perpendicular full switching of chiral antiferromagnetic order by current. *Nature* **607**, 474–479 (2022).

29. Pal, B. *et al.* Setting of the magnetic structure of chiral kagome antiferromagnets by a seeded spin-orbit torque. *Sci. Adv.* **8**, eabo5930 (2022).

30. Krishnaswamy, G. K. *et al.* Time-Dependent Multistate Switching of Topological Antiferromagnetic Order in $Mn_3Sn$. *Phys. Rev. Appl.* **18**, 024064 (2022).

31. Higo, T. & Nakatsuji, S. Thin film properties of the non-collinear Weyl antiferromagnet $Mn_3Sn$. *J. Magn. Magn. Mater.* **564**, 170176 (2022).

32. Kobayashi, Y., Shiota, Y., Narita, H., Ono, T. & Moriyama, T. Pulse-width dependence of spin–orbit torque switching in $Mn_3Sn$/Pt thin films. *Appl. Phys. Lett.* **122**, 122405 (2023).

33. Yoon, J.-Y. *et al.* Handedness anomaly in a non-collinear antiferromagnet under spin–orbit torque. *Nat. Mater.* **22**, 1106–1113 (2023).

34. Miron, I. M. *et al.* Perpendicular switching of a single ferromagnetic layer induced by in-plane current injection. *Nature* **476**, 189–193 (2011).

35. Liu, L., Lee, O. J., Gudmundsen, T. J., Ralph, D. C. & Buhrman, R. A. Current-Induced Switching of Perpendicularly Magnetized Magnetic Layers Using Spin Torque from the Spin Hall Effect. *Phys. Rev. Lett.* **109**, 096602 (2012).

36. Dieny, B. & Chshiev, M. Perpendicular magnetic anisotropy at transition metal/oxide interfaces and applications. *Rev. Mod. Phys.* **89**, 025008 (2017).

37. Higo, T. *et al.* Omnidirectional Control of Large Electrical Output in a Topological Antiferromagnet. *Adv. Funct. Mater.* **31**, 2008971 (2021).

38. Abo, G. S. *et al.* Definition of Magnetic Exchange Length. *IEEE Trans. Magn.* **49**, 4937–4939 (2013).

39. Frei, E. H., Shtrikman, S. & Treves, D. Critical Size and Nucleation Field of Ideal Ferromagnetic Particles. *Phys. Rev.* **106**, 446–455 (1957).

40. Attané, J. P., Ravelosona, D., Marty, A., Samson, Y. & Chappert, C. Thermally Activated Depinning of a Narrow Domain Wall from a Single Defect. *Phys. Rev. Lett.* **96**, 147204 (2006).

41. Mihai, A. P. *et al.* Magnetization reversal dominated by domain wall pinning in FePt based spin valves. *Appl. Phys. Lett.* **94**, 122509 (2009).

42. Burrowes, C. *et al.* Non-adiabatic spin-torques in narrow magnetic domain walls. *Nat. Phys.* **6**, 17–21 (2010).

43. Xie, H. *et al.* Magnetization switching in polycrystalline $Mn_3Sn$ thin film induced by self-generated spin-polarized current. *Nat. Commun.* **13**, 5744 (2022).

44. Zhu, L. Switching of Perpendicular Magnetization by Spin–Orbit Torque. *Adv. Mater.* **35**, 2300853 (2023).

45. Ikhlas, M. *et al.* Piezomagnetic switching of the anomalous Hall effect in an antiferromagnet at room temperature. *Nat. Phys.* **18**, 1086–1093 (2022).

46. Chen, X. *et al.* Octupole-driven magnetoresistance in an antiferromagnetic tunnel junction. *Nature* **613**, 490–495 (2023).

47. Matsuo, T., Higo, T., Nishio-Hamane, D. & Nakatsuji, S. Anomalous Hall effect in nanoscale structures of the antiferromagnetic Weyl semimetal $Mn_3Sn$ at room temperature. *Appl. Phys. Lett.* **121**, 013103 (2022).

48. Sato, Y. *et al.* Thermal stability of non-collinear antiferromagnetic $Mn_3Sn$ nanodot. *Appl. Phys. Lett.* **122**, 122404 (2023).

49. Fujita, H. Field-free, spin-current control of magnetization in non-collinear chiral antiferromagnets. *Phys. Status Solidi Rapid Res. Lett.* **11**, 1600360 (2017).

50. Nomoto, T. & Arita, R. Cluster multipole dynamics in noncollinear antiferromagnets. *Phys. Rev. Res.* **2**, 012045 (2020).

51. Xu, Z. *et al.* Deterministic spin-orbit torque switching including the interplay between spin polarization and kagome plane in $Mn_3Sn$. *Phys. Rev. B* **109**, 134433 (2024).

52. Suzuki, M.-T., Koretsune, T., Ochi, M. & Arita, R. Cluster multipole theory for anomalous Hall effect in antiferromagnets. *Phys. Rev. B* **95**, 094406 (2017).

53. Ikhlas, M. *et al.* Large anomalous Nernst effect at room temperature in a chiral antiferromagnet. *Nat. Phys.* **13**, 1085–1090 (2017).

54. Li, X. *et al.* Anomalous Nernst and Righi-Leduc Effects in $Mn_3Sn$ : Berry Curvature and Entropy Flow. *Phys. Rev. Lett.* **119**, 056601 (2017).

55. Higo, T. *et al.* Large magneto-optical Kerr effect and imaging of magnetic octupole domains in an antiferromagnetic metal. *Nat. Photonics* **12**, 73–78 (2018).

56. Tomiyoshi, S. & Yamaguchi, Y. Magnetic Structure and Weak Ferromagnetism of $Mn_3Sn$ Studied by Polarized Neutron Diffraction. *J. Phys. Soc. Jpn.* **51**, 2478–2486 (1982).

57. Brown, P. J., Nunez, V., Tasset, F., Forsyth, J. B. & Radhakrishna, P. Determination of the magnetic structure of $Mn_3Sn$ using generalized neutron polarization analysis. *J. Phys.: Condens. Matter* **2**, 9409–9422 (1990).

58. Yang, H. *et al.* Topological Weyl semimetals in the chiral antiferromagnetic materials $Mn_3Ge$ and $Mn_3Sn$. *New J. Phys.* **19**, 015008 (2017).

59. Tomiyoshi, S. Polarized Neutron Diffraction Study of the Spin Structure of $Mn_3Sn$. *J. Phys. Soc. Jpn.* **51**, 803–810 (1982).

60. Cable, J. W., Wakabayashi, N. & Radhakrishna, P. A neutron study of the magnetic structure of $Mn_3Sn$. *Solid State Commun.* **88**, 161–166 (1993).

61. Duan, T. F. *et al.* Magnetic anisotropy of single-crystalline $Mn_3Sn$ in triangular and helix-phase states. *Appl. Phys. Lett.* **107**, 082403 (2015).

62. Sung, N. H., Ronning, F., Thompson, J. D. & Bauer, E. D. Magnetic phase dependence of the anomalous Hall effect in $Mn_3Sn$ single crystals. *Appl. Phys. Lett.* **112**, 132406 (2018).

63. Shukla, A., Qian, S. & Rakheja, S. Impact of strain on the SOT-driven dynamics of thin film $Mn_3Sn$. *J. Appl. Phys.* **135**, 123903 (2024).

64. Zheng, Z. *et al.* Effective electrical manipulation of a topological antiferromagnet by orbital torques. *Nat. Commun.* **15**, 745 (2024).

65. Lee, K.-S., Lee, S.-W., Min, B.-C. & Lee, K.-J. Threshold current for switching of a perpendicular magnetic layer induced by spin Hall effect. *Appl. Phys. Lett.* **102**, 112410 (2013).

66. Roschewsky, N., Lambert, C.-H. & Salahuddin, S. Spin-orbit torque switching of ultralarge-thickness ferrimagnetic GdFeCo. *Phys. Rev. B* **96**, 064406 (2017).

67. Krén, E., Paitz, J., Zimmer, G. & Zsoldos, É. Study of the magnetic phase transformation in the $Mn_3Sn$ phase. *Physica B+C* **80**, 226–230 (1975).

68. Ikeda, T. *et al.* Improvement of Large Anomalous Hall Effect in Polycrystalline Antiferromagnetic $Mn_{3+x}Sn$ Thin Films. *IEEE Trans. Magn.* **55**, 1–4 (2019).

69. Meng, Q. *et al.* Magnetostriction, piezomagnetism and domain nucleation in a Kagome antiferromagnet. *Nat. Commun.* **15**, 6921 (2024).

70. Deng, Y., Li, R. & Liu, X. Thickness dependent anomalous Hall effect in noncollinear antiferromagnetic $Mn_3Sn$ polycrystalline thin films. *J. Alloys Compd.* **874**, 159910 (2021).

71. Li, S. *et al.* Nanoscale Magnetic Domains in Polycrystalline $Mn_3Sn$ Films Imaged by a Scanning Single-Spin Magnetometer. *Nano Lett.* **23**, 5326–5333 (2023).

72. Fukami, S., Zhang, C., DuttaGupta, S., Kurenkov, A. & Ohno, H. Magnetization switching by spin–orbit torque in an antiferromagnet–ferromagnet bilayer system. *Nat. Mater.* **15**, 535–541 (2016).

73. Zhang, R. Q. *et al.* Current-induced magnetization switching in a CoTb amorphous single layer. *Phys. Rev. B* **101**, 214418 (2020).

74. Zheng, Z. *et al.* Field-free spin-orbit torque-induced switching of perpendicular magnetization in a ferrimagnetic layer with a vertical composition gradient. *Nat. Commun.* **12**, 4555 (2021).

75. Takeuchi, Y. *et al.* Electrical coherent driving of chiral antiferromagnet. *Science* **389**, 830–834 (2025).

76. Nihei, K. *et al.* Temperature dependence of current-induced switching in thin epitaxial films of Mn3Sn: Revealing the dominant role of spin–orbit torque. *Appl. Phys. Lett.* **127**, 132406 (2025).

77. Momma, K. & Izumi, F. *VESTA* : a three-dimensional visualization system for electronic and structural analysis. *J. Appl. Crystallogr.* **41**, 653–658 (2008).

78. Cook, H. C. Investigation of Sputtered β-Tantalum Thin Films. *J. Vac. Sci. Technol.* **4**, 80–86 (1967).

79. Canny, J. A Computational Approach to Edge Detection. *IEEE Trans. Pattern Anal. Mach. Intell.* **PAMI-8**, 679–698 (1986).

Supplementary Material for

# Crossover between intrinsic and temperature-assisted regimes in spin-orbit torque switching of antiferromagnetic order

Takumi Matsuo[1,2,†], Tomoya Higo[1,3,4†,*], Hanshen Tsai[1], Daisuke Nishio-Hamane[4], Takuya Matsuda[1,5], Ryota Uesugi[1,4], Kouta Kondou[6,7], Shinji Miwa[4], Yoshichika Otani[4,6], and Satoru Nakatsuji[1,2,4,8,9*]

*[1]Department of Physics, The University of Tokyo, Bunkyo-ku, Tokyo 113-0033, Japan*

*[2]Department of Physics and Astronomy, The Johns Hopkins University, Baltimore, Maryland 21218, USA*

*[3]PRESTO, Japan Science and Technology Agency, Kawaguchi, Saitama 332-0012, Japan*

*[4]Institute for Solid State Physics, The University of Tokyo, Kashiwa, Chiba 277-8581, Japan.*

*[5]Graduate School of Engineering Science, Osaka University, Toyonaka, Osaka 560-8531, Japan*

*[6]Center for Emergent Matter Science, RIKEN, Wako, Saitama 351-0198, Japan*

*[7]Institute for Open and Transdisciplinary Research Initiatives, Osaka University, Toyonaka, Osaka 560-8531, Japan*

*[8]Trans-scale Quantum Science Institute, The University of Tokyo, Bunkyo-ku, Tokyo 113-0033, Japan*

*[9]Canadian Institute for Advanced Research, Toronto, Ontario M5G 1Z7, Canada*

## SM1. Optimization of the annealing temperature

The annealing temperature $T_a$ of our multilayers was chosen to optimize several key qualities in our films: namely, the absence of impurities, the surface roughness, and the anomalous Hall conductivity.

Fig. S1 shows the X-ray diffraction (XRD) spectra obtained by $2\theta/\omega$ scans for $Mn_3Sn$(40 nm)/Ta(5 nm)/$AlO_x$(3 nm) films annealed at different $T_a$ after room temperature deposition of all layers. When 400°C $\leq T_a \leq$ 600°C, all peaks can be attributed to $Mn_3Sn$ or the substrate, indicating that the reaction between $Mn_3Sn$ and Ta is negligible. For $T_a$=700°C, a peak unassociated with $Mn_3Sn$ emerges near 29°. We thus conclude that our $Mn_3Sn$ layers are of single phase when $T_a$<700°C.

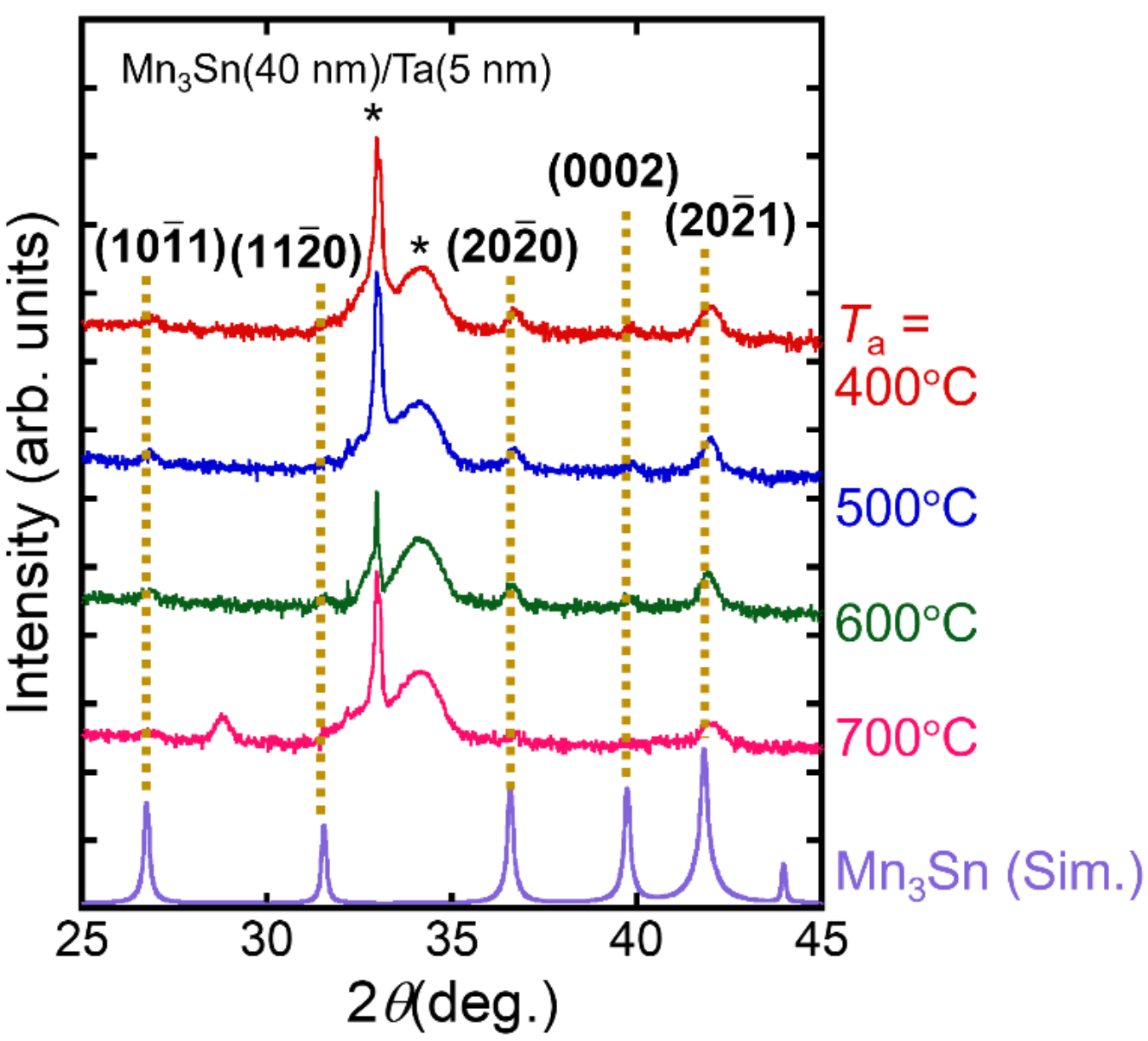

**Fig. S1.** XRD $2\theta/\omega$ spectra of $Mn_3Sn$(40 nm)/Ta(5 nm)/$AlO_x$(3 nm) films annealed at various temperatures $T_a$. The simulated spectrum of a randomly oriented $Mn_3Sn$ sample is shown at the bottom. The asterisks indicate the substrate peaks.

Atomic force microscopy (AFM) was employed to characterize the surface roughness of our $Mn_3Sn$/Ta films. Fig. S2**a** shows AFM images of CA films with various annealing temperatures $T_a$ and the root mean square (RMS) roughness calculated from these images is plotted in Fig. S2**b**. When $T_a \leq 600$°C, i.e., when the $Mn_3Sn$ and Ta layers have not reacted, the surface roughness increases slightly with $T_a$ from ~0.4 nm to ~0.6 nm. These values are significantly smaller than those of previously reported polycrystalline $Mn_3Sn$ films, where thermal processing was conducted without capping[1]. At $T_a$=700°C, there is a large jump in the RMS value to 1.4 nm, indicating that the reaction between $Mn_3Sn$ and Ta deforms the film structure. The RMS value for the $T_a$ =500°C sample measured by AFM matches the $Mn_3Sn$/Ta interfacial roughness obtained by transmission electron microscopy (TEM) (Fig. 2**c** in the main text), indicating that the surface roughness we obtain from AFM corresponds to the $Mn_3Sn$/Ta interfacial quality.

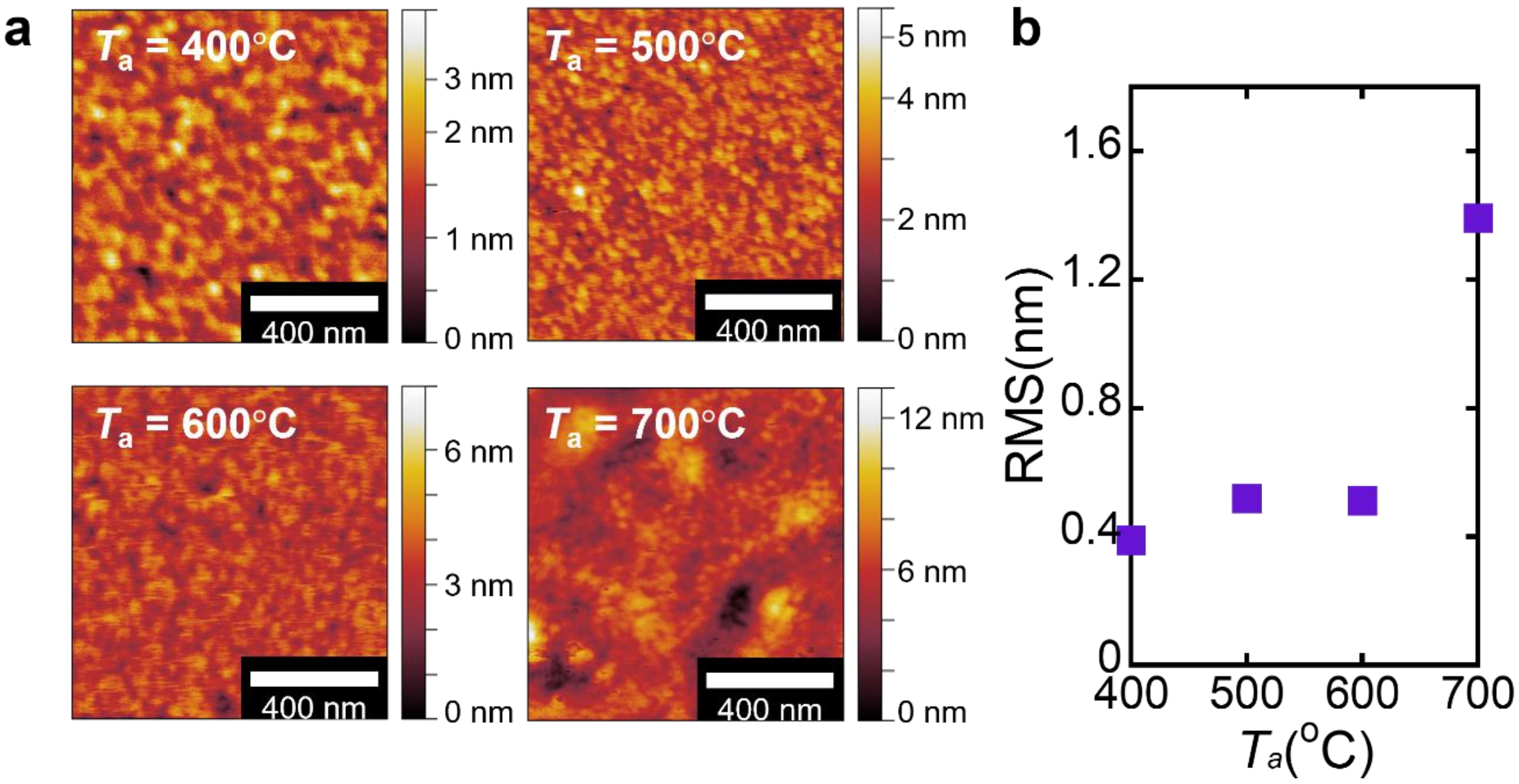

**Fig. S2.** **a**, Atomic force microscopy (AFM) images of $Mn_3Sn$(40 nm)/Ta(5 nm)/$AlO_x$(3 nm) films grown with various $T_\mathrm{a}$. The image for $T_\mathrm{a}$ =500°C is identical to Fig. 2**b** in the main text. **b**, RMS values calculated from the AFM images as a function of $T_\mathrm{a}$.

Finally, the anomalous Hall effect (AHE) was measured under a perpendicular magnetic field ($H$) sweep at various temperatures for films ($t$ = 40 nm) with different $T_a$ ranging from 400°C to 700°C. To evaluate the Hall resistivity and Hall conductivity of the $Mn_3Sn$ after taking into account shunting effects into the Ta layer, we employed the parallel resistor model. In this model, The $Mn_3Sn$ and Ta layers are each modeled to be resistors that are connected in parallel with each other. Each layer is modeled to have a resistance of $R_i = \rho_i l/wt_i$ ($i$=$Mn_3Sn$, Ta), where $\rho_i$ is the longitudinal resistivity of the respective material, $l$ and $w$ are the channel length and width, and $t_i$ is the thickness of the respective layer. Note that $t_{\mathrm{Mn_3Sn}}$ corresponds to what is defined as $t$ in the main text. The ratio of the current flowing in the $Mn_3Sn$ and Ta layers can be estimated as $I_{\mathrm{Mn_3Sn}}\colon I_{\mathrm{Ta}} = R_{\mathrm{Ta}}\colon R_{\mathrm{Mn_3Sn}}$. Therefore, one can calculate the current flowing in the $Mn_3Sn$ layer as

$$I_{\mathrm{Mn_3Sn}} = I\frac{R_{\mathrm{Ta}}}{R_{\mathrm{Mn_3Sn}}\, R_{\mathrm{Ta}}} = \frac{I}{1+\dfrac{\rho_{\mathrm{Mn_3Sn}}/t_{\mathrm{Mn_3Sn}}}{\rho_{\mathrm{Ta}}/t_{\mathrm{Ta}}}}$$

Furthermore, the anomalous Hall effect from the $Mn_3Sn$ layer drives a transverse current in the Ta layer as well as in the $Mn_3Sn$ layer because the two layers are in electrical contact with each other. The Hall voltage measured in the bilayer device therefore can be written as

$$V_{yx} = V_{yx,\mathrm{Mn_3Sn}} \frac{R_{\mathrm{Ta}}}{R_{\mathrm{Mn_3Sn}}\, R_{\mathrm{Ta}}} = \frac{V_{yx,\mathrm{Mn_3Sn}}}{1 + \dfrac{\rho_{\mathrm{Mn_3Sn}}/t_{\mathrm{Mn_3Sn}}}{\rho_{\mathrm{Ta}}/t_{\mathrm{Ta}}}}$$

Using these relations, the transverse (Hall) resistivity and transverse (Hall) conductivity of the $Mn_3Sn$ layer, $\rho_{yx,\mathrm{Mn_3Sn}}$ can be estimated as

$$\rho_{yx,\mathrm{Mn_3Sn}} = t_{\mathrm{Mn_3Sn}} \frac{V_{yx,\mathrm{Mn_3Sn}}}{I_{\mathrm{Mn_3Sn}}} = t_{\mathrm{Mn_3Sn}} \frac{V_{yx}}{I} \left( 1 + \frac{\dfrac{\rho_{\mathrm{Mn_3Sn}}}{t_{\mathrm{Mn_3Sn}}}}{\dfrac{\rho_{\mathrm{Ta}}}{t_{\mathrm{Ta}}}} \right)^2$$

using which the Hall conductivity $\sigma_{yx,\mathrm{Mn_3Sn}} = -\frac{\rho_{yx,\mathrm{Mn_3Sn}}}{\rho^2_{\mathrm{Mn_3Sn}} + \rho^2_{yx,\mathrm{Mn_3Sn}}}$ can also be calculated.

The temperature dependence of the spontaneous Hall conductivities $\sigma_{yx}(H = 0)$ ($= \frac{\sigma_{yx}(\mu_0 H = -0) - \sigma_{yx}(\mu_0 H = +0)}{2}$, where $\mu_0 H = +0$ indicates $\mu_0 H = 0$ T while sweeping the magnetic field from positive to negative and vice versa) for these films is shown in Fig. S3. The spontaneous Hall signal is intact at all temperatures for films with $T_\mathrm{a} \leq 500$°C. At $T_\mathrm{a} = 600$°C, however, the spontaneous Hall response vanishes below 200 K, leaving only an ordinary Hall effect. This change in behavior closely resembles the results reported by Ikeda *et al.*[2] where the composition dependence of the anomalous Hall response was investigated. In their study, $Mn_{78}Sn_{22}$ thin films were shown to retain a finite hysteresis at low temperatures down to 50 K with a minimum around 100 K, while the hysteresis in $Mn_{77}Sn_{23}$ films vanished below 200 K. The disappearance of the AHE in the latter temperature range has been attributed to a transition from the inverse triangular structure to a spiral spin structure[3–5]. Comparing our results with this study, we can surmise that the local Mn-Sn composition is affected by $T_\mathrm{a}$: at higher $T_\mathrm{a}$, Mn atoms may mobilize towards the crystallite boundaries, not contributing to the characteristic responses of $Mn_3Sn$. For the film with $T_\mathrm{a} = 700$°C, a change in the sign of $\sigma_{yx}(H = 0)$ is observed below 200 K, presumably due to the impurity detected in the XRD spectrum (Fig. S1).

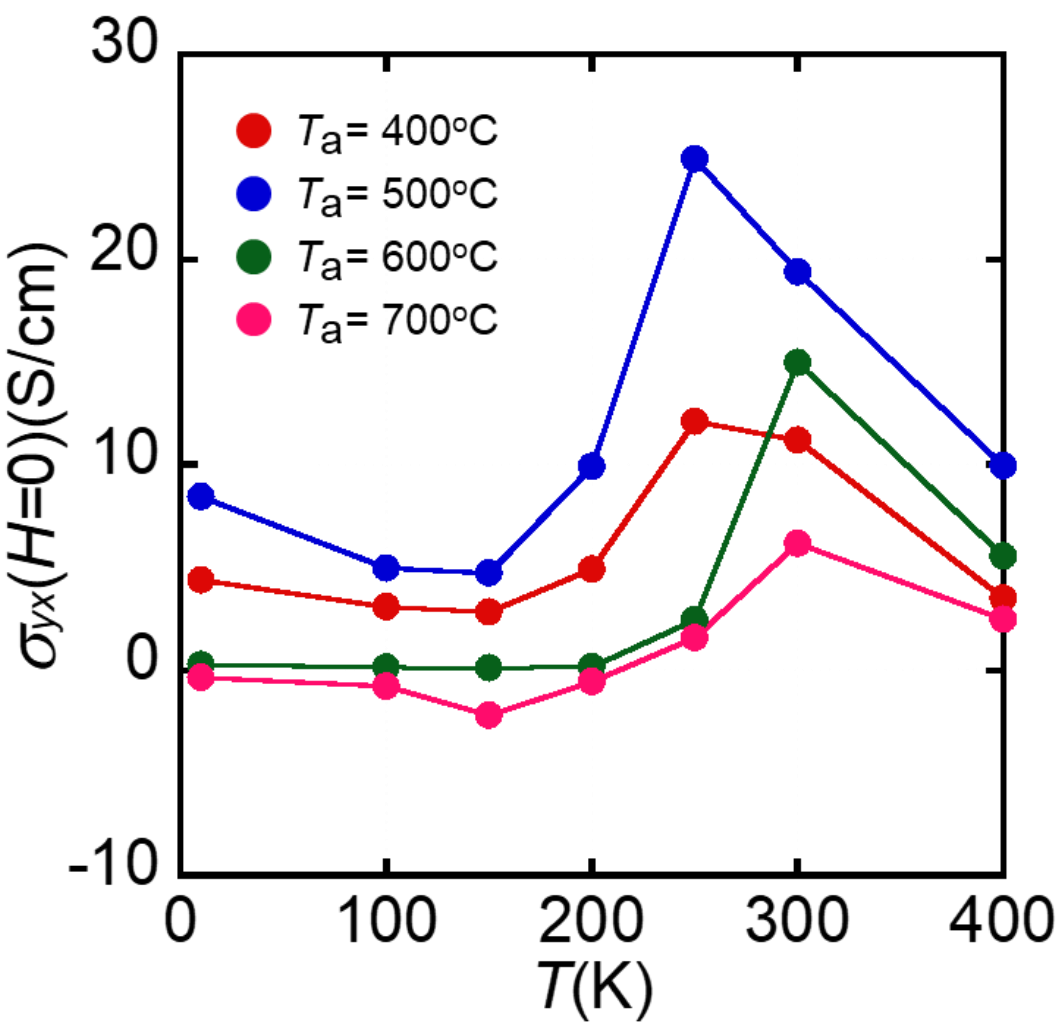

**Fig. S3.** Temperature dependence of the spontaneous anomalous Hall conductivity $\sigma_{yx}(H = 0)$ for $Mn_3Sn$(40 nm)/Ta(5 nm)/$AlO_x$(3 nm) annealed at various temperatures $T_a$. The shaded region indicates the temperature range in which the inverse triangular structure is expected to be intact.

**SM2**: Additional analysis of the $Mn_3Sn$/Ta interface

Fig. S4 shows the results of our X-ray reflectivity (XRR) measurements on our $Mn_3Sn$(40 nm)/Ta(5 nm)/$AlO_x$(3 nm) film. The roughness of all interfaces extracted from the XRR fit are less than 0.7 nm, confirming, in addition to the AFM and TEM data in our main text, the superior interfacial properties of our heterostructure.

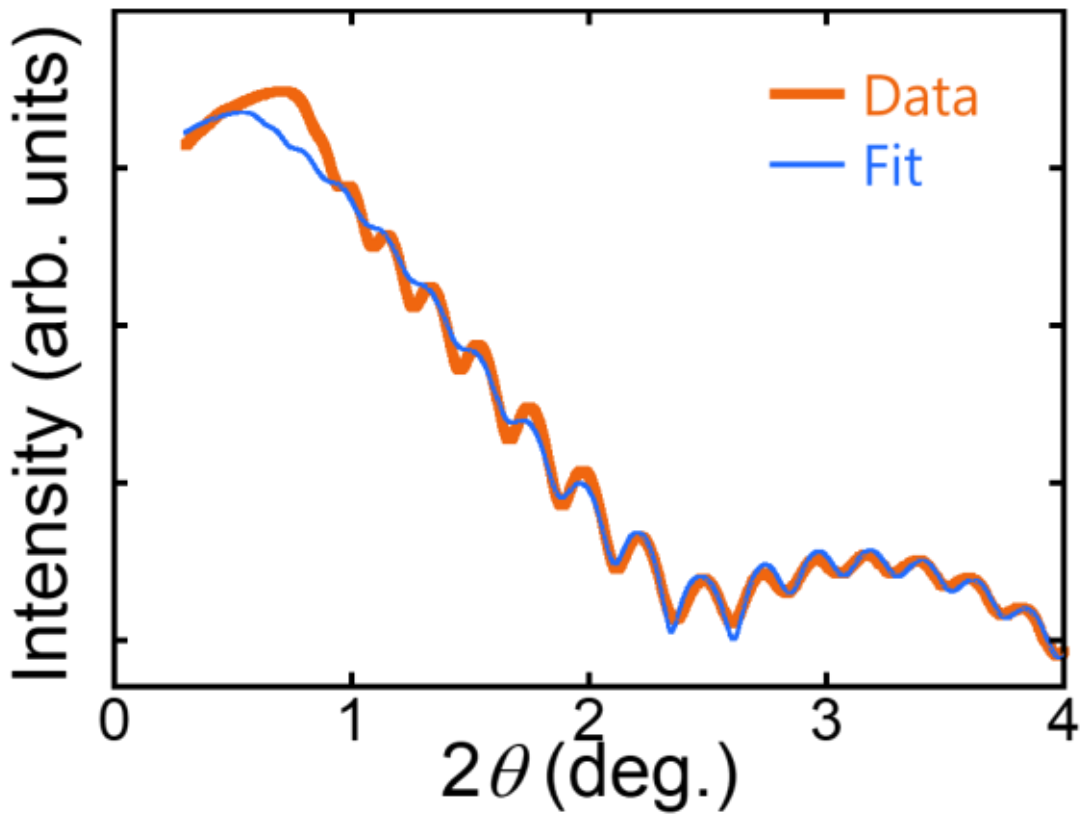

**Fig. S4.** X-ray reflectivity results on a $Mn_3Sn$(40 nm)/Ta(5 nm)/$AlO_x$(3 nm) film. The thin blue line represents the fit to the measured data.

To characterize the interdiffusion of the $Mn_3Sn$ and Ta layers in detail, we map the elemental distribution measured by energy dispersive X-ray spectroscopy (EDX) in the cross section of our $Mn_3Sn$($t$=15 nm)/Ta(5 nm)/$AlO_x$(3 nm) film using scanning transmission electron microscopy (STEM) as shown in Fig. S5. We see that the vast majority of interdiffusion between the $Mn_3Sn$ and Ta layers is contained within a nanometer of the interface, a negligible portion of the $Mn_3Sn$ layers with thicknesses from 15 to 200 nm we employ in this study.

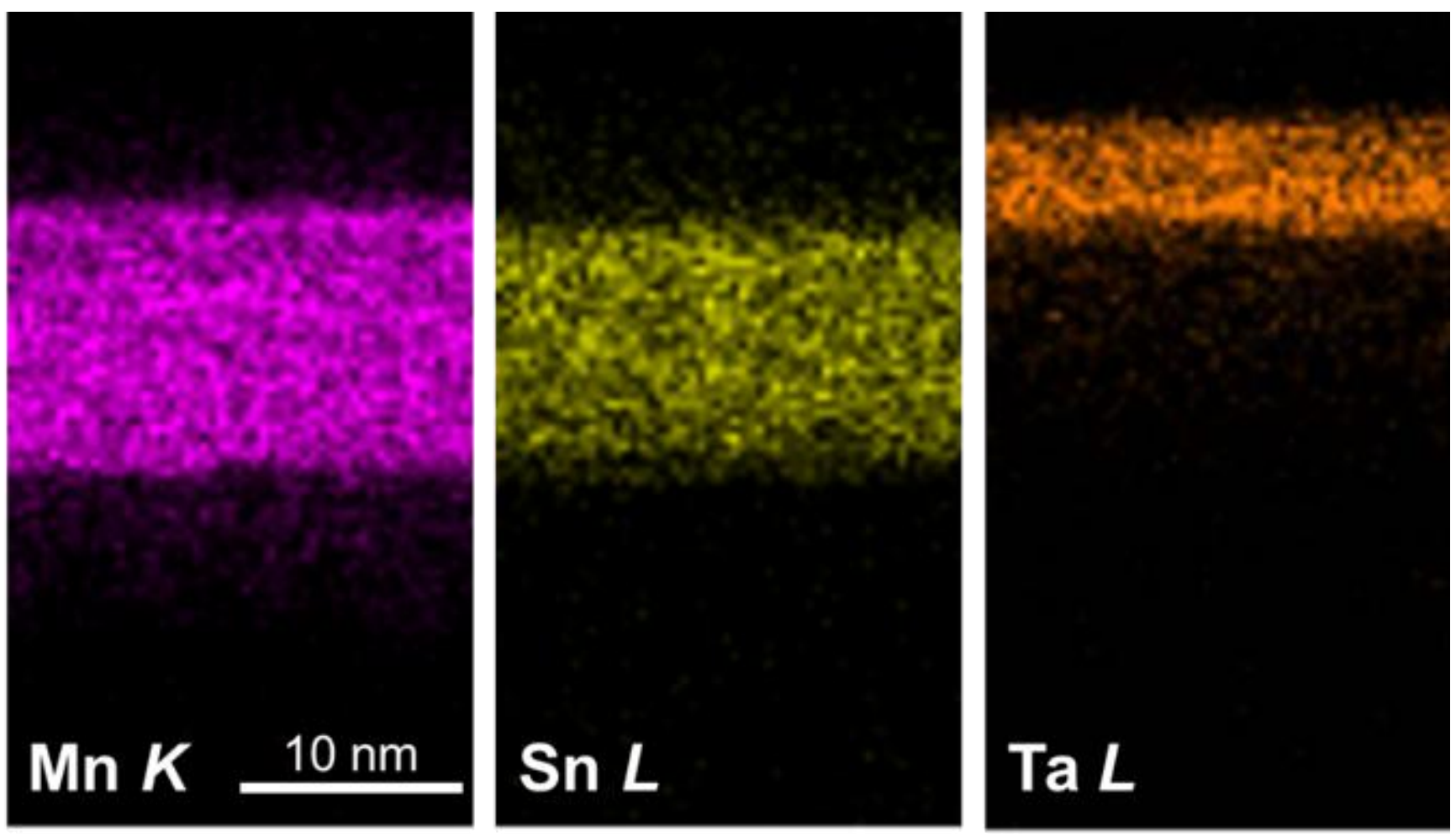

**Fig. S5.** Elemental profiles of a $Mn_3Sn$(15 nm)/Ta(5 nm)/$AlO_x$(3 nm) film measured by STEM-EDX.

## SM3. Characterization of the crystal domain size

In Fig. S6, we show a high-resolution TEM (HRTEM) image of our $t$ = 15 nm film. We observe a crystalline region with a width of roughly 40 nm (with boundaries indicated by the yellow lines), indicating that the lateral dimensions of the crystal grains in our films are $<$ ~100 nm. We thus assume that the Hall bars, which have dimensions of tens of microns, have on average the same crystal properties indicated by XRD on the bare film.

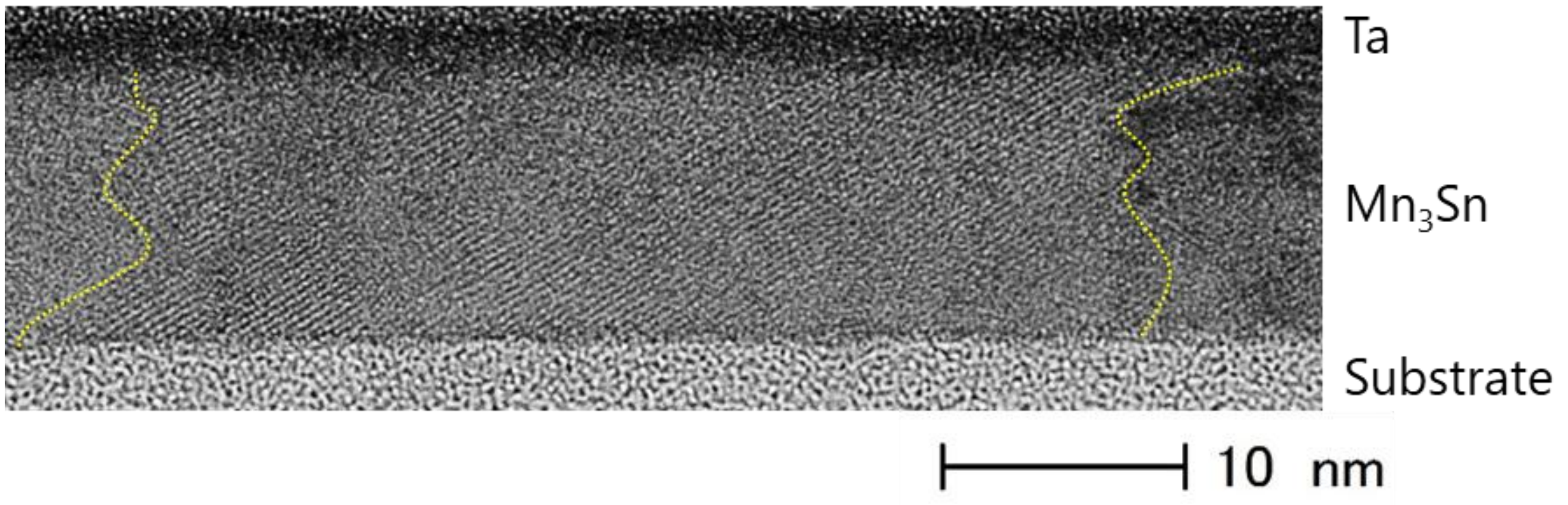

**Fig. S6.** Lateral size of crystal domains in the $Mn_3Sn$ layer indicated by HRTEM.

## SM4. In-plane $2\theta\chi/\phi$ X-ray diffraction scans for thinner films

One caveat of the $2\theta/\omega$ scans shown in Fig. 2**d** is that the signals from the thinner samples are much smaller due to the reduced sample volume, and comparison of the peak intensities becomes challenging. Therefore, to complement the $2\theta/\omega$ scans, in-plane $2\theta\chi/\phi$scans were performed for films with $t = 15$, 20, and 25 nm where the X-ray source and detector were placed such that the scattering vector is perpendicular to the sample normal. This geometry circumvents the restriction of the signal size stemming from the small sample volumes. The results of these measurements are shown in Fig. S7**a**. Note that while the $2\theta/\omega$ scans are sensitive to the out-of-film plane crystal planes, the $2\theta\chi/\phi$scans pick up the crystal planes that lie in the film plane. The intensity of the (0002) peaks are significantly larger compared to what is expected from a powder sample with completely random orientation (Fig. S7**b**), confirming that the Kagome planes lie perpendicular to the film plane. We thus confirm the preferentially perpendicular Kagome plane orientation throughout the employed thickness range.

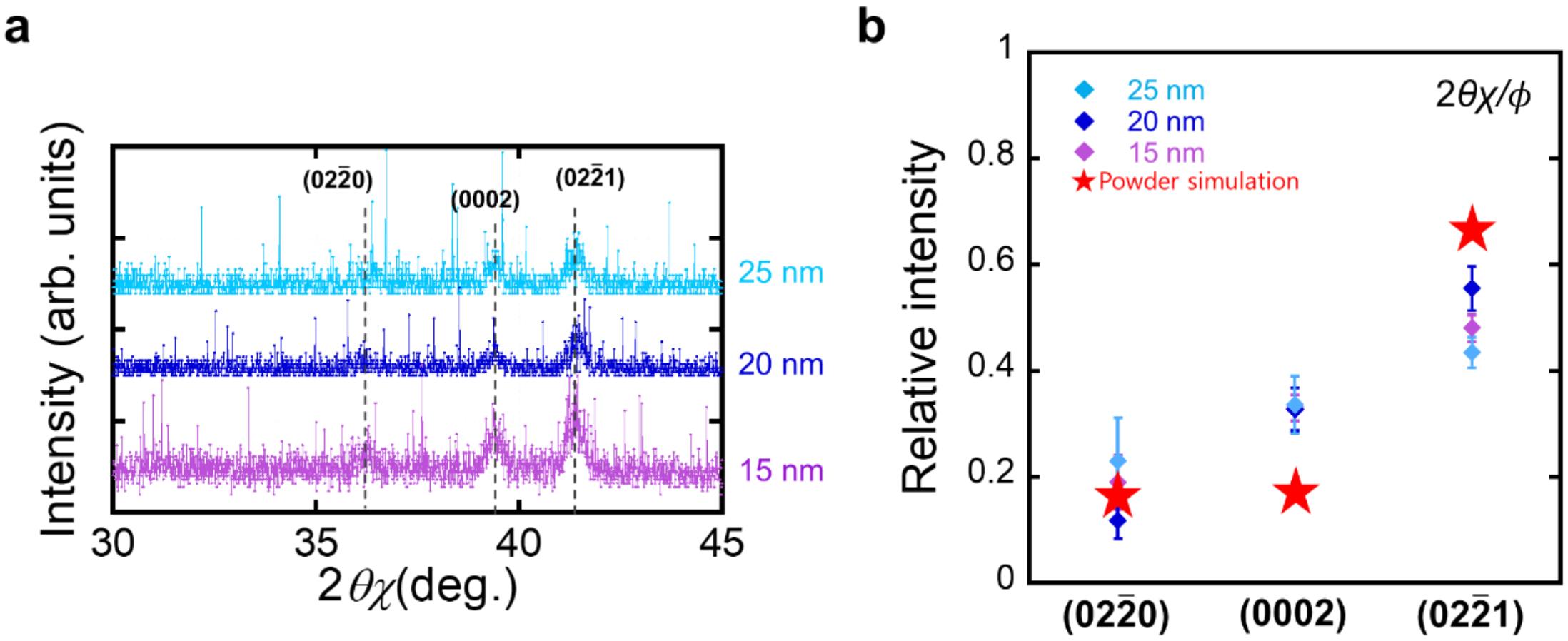

**Fig. S7. a,** In-plane $2\theta\chi/\phi$ X-ray diffraction signals measured for $Mn_3Sn(t)/Ta(5 nm)/AlO_x(3 nm)$ films with 15 nm $\leq t \leq$ 25 nm ($T_a$ = 500°C). The indicated peak positions and corresponding orientations are those expected from randomly oriented $Mn_3Sn$ (shown in Fig. 2**d** of the main text). **b,** Relative peak intensities for the (02$\bar{2}$0), (0002), and (02$\bar{2}$1) peaks.

## SM5: Influence of the presence of the capping layer during annealing on the film crystallinity

To elucidate the effects of the capping Ta and $AlO_x$ layers during the annealing process of $Mn_3Sn$, we compare the XRD scans of $Mn_3Sn(40 nm)/Ta(5 nm)/AlO_x(3 nm)$ films where all layers were annealed together at 500°C versus a reference film where the $Mn_3Sn$ was annealed at 500°C before the growth of the Ta and $AlO_x$ layers. All deposition conditions besides the annealing order, including the sputtering chamber, were unchanged. The $2\theta/\omega$ XRD spectra of both films, along with simulated data assuming a completely random crystal orientation, are shown in Fig. S8. Most notably, the peak corresponding to (0002) is subdued when the $Mn_3Sn$ layer is annealed with capping, while the spectrum closely resembles that of a random orientation when the annealing is conducted without capping. This suggests that the preferentially out-of-plane orientation of the Kagome planes in our samples established in Fig. 2**d** of the main text and SM4 is a direct effect of annealing the $Mn_3Sn$ with capping layers.

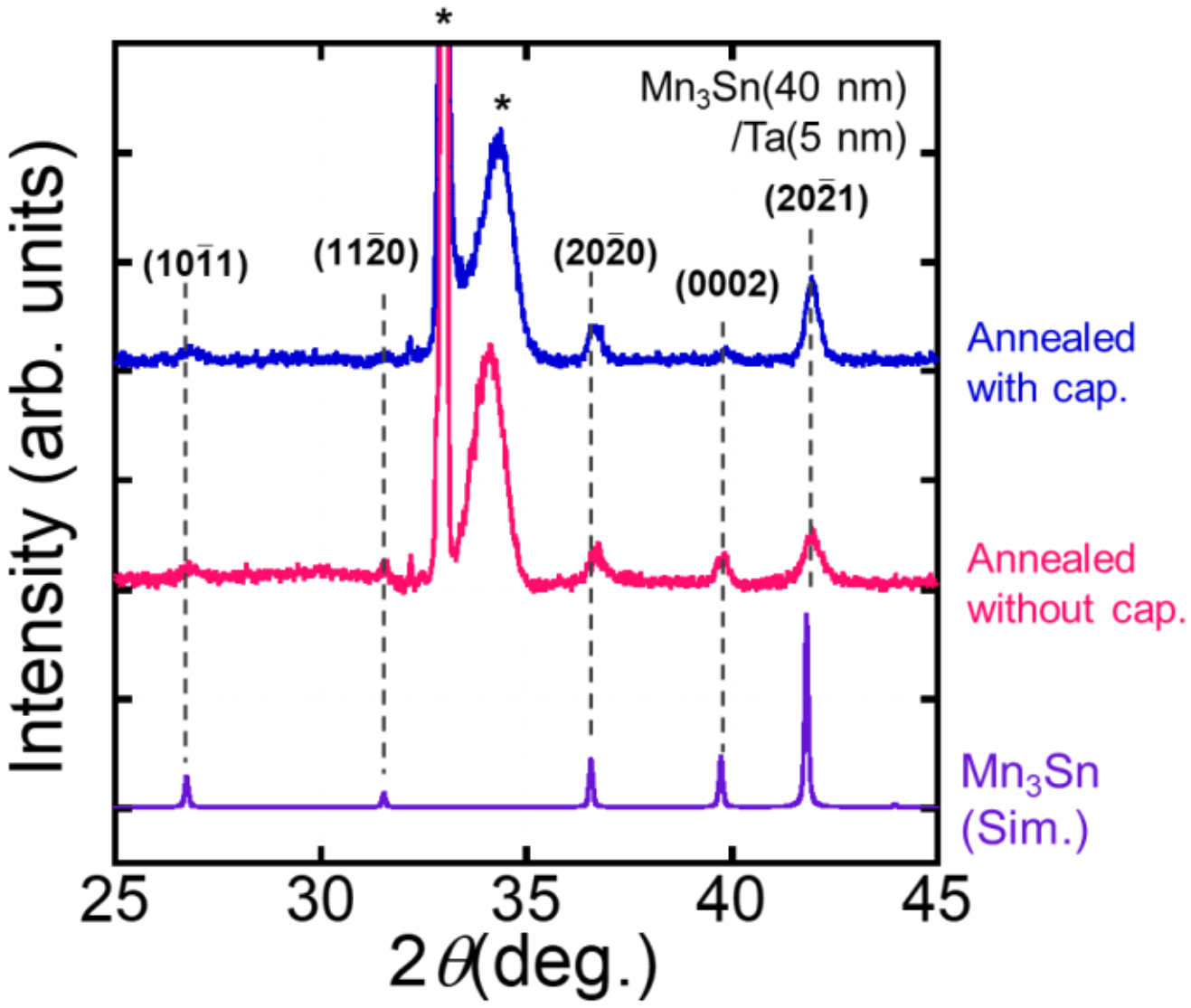

**Fig. S8.** XRD signals of $Mn_3Sn$(40 nm)/Ta(5 nm)/$AlO_x$(3 nm) films annealed at 500°C before (blue) and after (pink) deposition of the Ta and $AlO_x$ layers. The simulated pattern for randomly oriented $Mn_3Sn$ is shown at the bottom. The asterisks represent the substrate peaks.

**SM6. Hall conductivities of devices**

The field dependencies of the Hall conductivities as a function of out-of-film plane magnetic field of the representative $t$=20 nm ($<t_c$) and $t$=60 nm($>t_c$) devices used in the switching measurements are shown in Fig. S9. The spontaneous Hall conductivities are comparable with previous reports on both bulk single crystal[6] and thin film[7] $Mn_3Sn$ which have concluded that the predominant contribution to the anomalous Hall effect is the Berry curvature-driven mechanism.

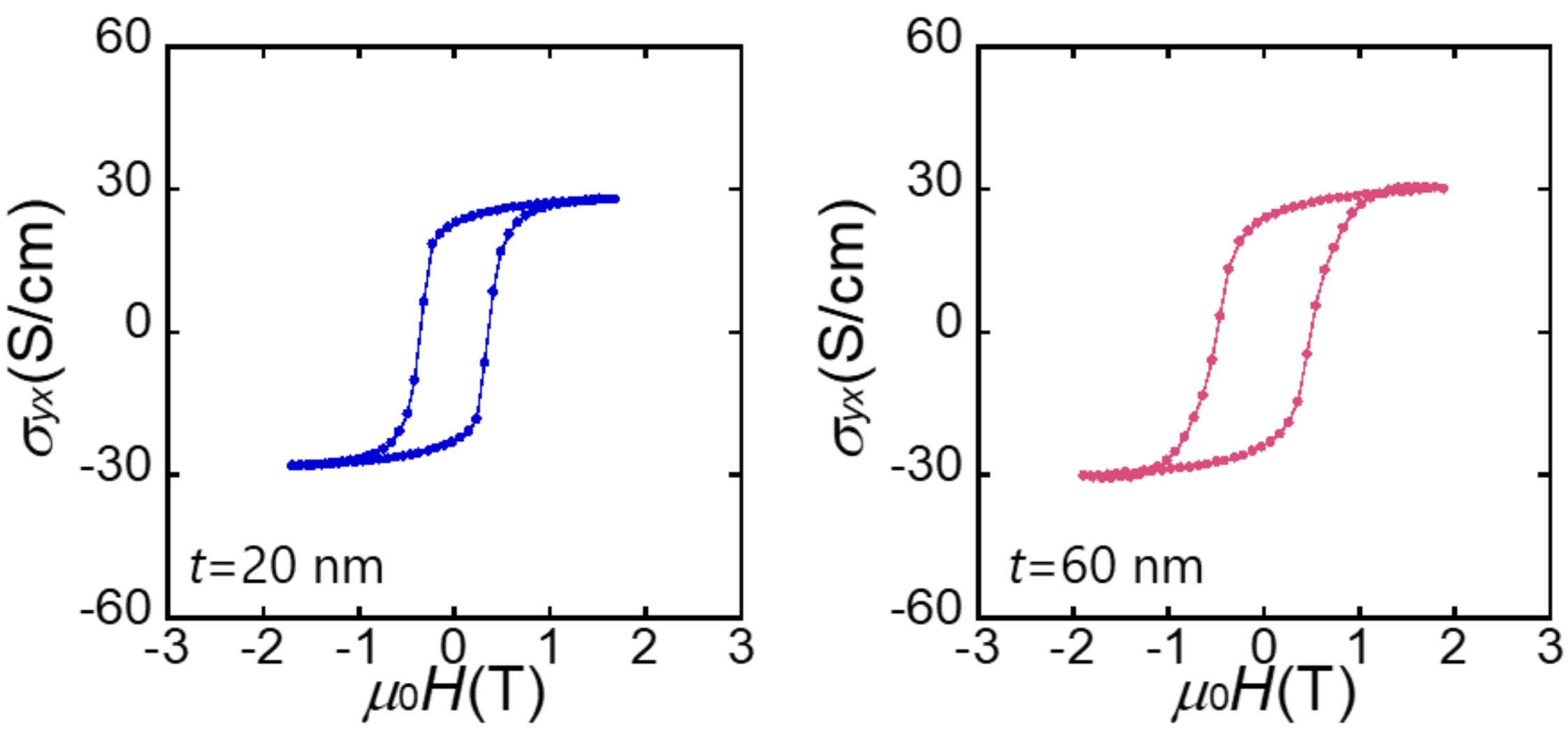

**Fig. S9.** Hall conductivity of the $Mn_3Sn$ layer in our $Mn_3Sn$($t$)/Ta(5 nm)/AlO$x$(3 nm) films for $t$=20 nm (left) and $t$=60 nm (right) as a function of out-of-plane magnetic field for all thicknesses.

**SM7. Overview of numerical simulations of device temperatures**

The "Electromagnetic Heating" module in the commercial software COMSOL Multiphysics was used to perform numerical investigations of the Joule heating in our devices. In the model, a Hall bar consisting of a $Mn_3Sn$($t$)/Ta(5 nm) bilayer with the same dimensions as those used in our measurements (channel width = 16 μm; channel length = 96 μm; see Fig. 3**a** of the main text) is placed on a Si(380 μm)/$SiO_2$(500 nm) substrate (Fig. S10). Au (200 nm) contact pads are placed on top of the Hall bars to replicate our experimental setup. The steady-state temperature distribution $T$ under a current in the $Mn_3Sn$/Ta bilayer is calculated by numerically solving the equation

$$0 = \nabla \cdot (k\nabla T) + \rho j^2$$

where $k$ is the thermal heat conductivity, $j$ is the electrical current density, and $\rho$ is the electrical resistivity. The thermal conductivities for $Mn_3Sn$ and Ta were taken from Refs.[8,9], while the values in the COMSOL material library were used for Au, Si, and $SiO_2$. $\rho_{\mathrm{Mn_3Sn}}$ was estimated through the parallel resistor model using the experimental longitudinal resistance of our devices and $\rho_{\mathrm{Ta}} = 200$ μΩ cm as the electrical resistivity for Ta[10]. All parameters were assumed to be independent of temperature.

We set the bottom of the Si substrate to have a fixed temperature of $T_0 = 293.15$ K to model the room temperature setup of our switching measurements. We further assumed heat transfer to the surrounding air (also set to be at $T_0 = 293.15$ K) from any exposed surfaces described by the relation

$$q = h\Delta T,$$

where $q$ is the heat flux, $h$ is the heat transfer coefficient, and $\Delta T$ is the difference in temperature between the device and surrounding air. $h$ was fixed to be 5 $W/m^2K$ (a typical value used to describe natural thermal convection to air[11]) unless otherwise specified. The temperature at the point located on the $Mn_3Sn$/Ta interface and in the center of the lateral Hall bar geometry was chosen as a characteristic value representing the device temperature.

The time evolution of the temperature distribution in the model shown in Fig. S10 was also calculated using the "Electromagnetic Heating" module in COMSOL Multiphysics, this time solving the time-dependent equation,

$$DC_p \frac{\partial T}{\partial \tau} = \nabla \cdot (k\nabla T) + \rho J^2,$$

where $D$ is the density, $C_p$ is the specific heat under constant pressure, and $\tau$ is time. The current is turned on instantaneously at $\tau = 0$ ns and is similarly instantaneously removed at $\tau = 20000$ ns; i.e. the simulated pulse width was set to be 20 μs, which, according to Refs.[12,13], is significantly longer than timescales typical of Joule heating in devices with similar structures and dimensions. The densities and specific heat values for $Mn_3Sn$ and Ta was taken from Refs.[3,9,14]. All other material properties and boundary conditions were shared with the steady-state simulations (including the input current). Again, all parameters were assumed to be independent of temperature. The results of the time-dependent calculations are shown in SM10.

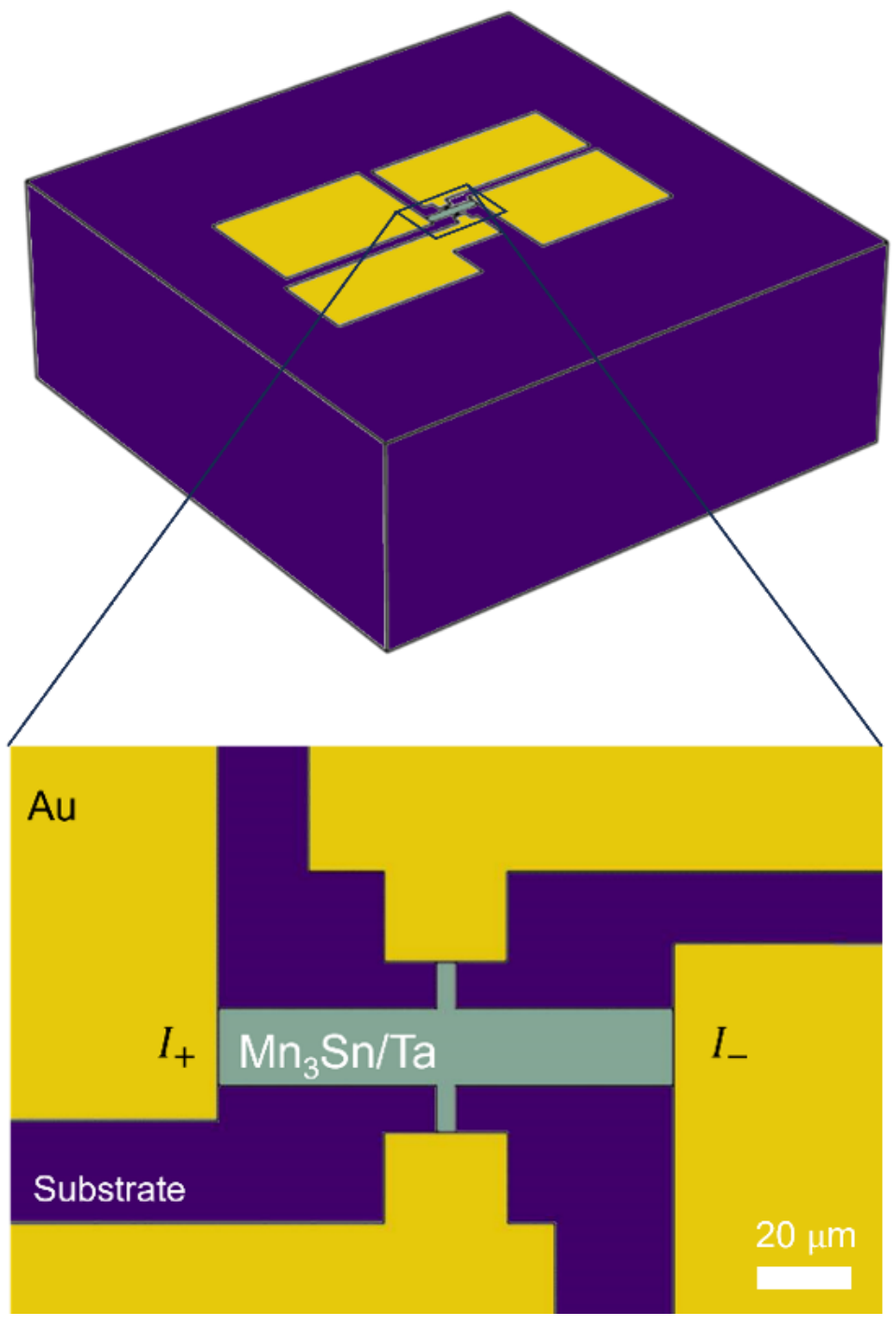

**Fig. S10,** Geometry used in COMSOL simulations of Joule heating. The substrate is modeled as a 500 nm thick $SiO_2$ layer on top of a 380 μm thick Si block. The Hall bar is modeled as $Mn_3Sn(t)$/Ta(5 nm) bilayer where $t$ is varied from 15 to 200 nm. The Au contact pads are modeled as 200 nm thick films. The channel width and length of the Hall bar is designed to be 16 μm and 96 μm, respectively.

**SM8: Contribution of heat transfer to surrounding air to device temperatures**

We show in Fig. S11 the calculated steady state temperature for devices of various $t$ upon application of the experimentally observed critical current using several values of $h$, the heat transfer coefficient describing thermal convection to the surrounding air (the results for $h =$ 5 W/m$^2$K are identical to those shown in Fig. 5**c** of the main text). Varying $h$ over two orders of magnitude had practically no effect on the obtained temperature value, indicating that heat transfer to the surrounding environment is negligible and most of the heat in the device due to Joule heating escapes through the substrate.

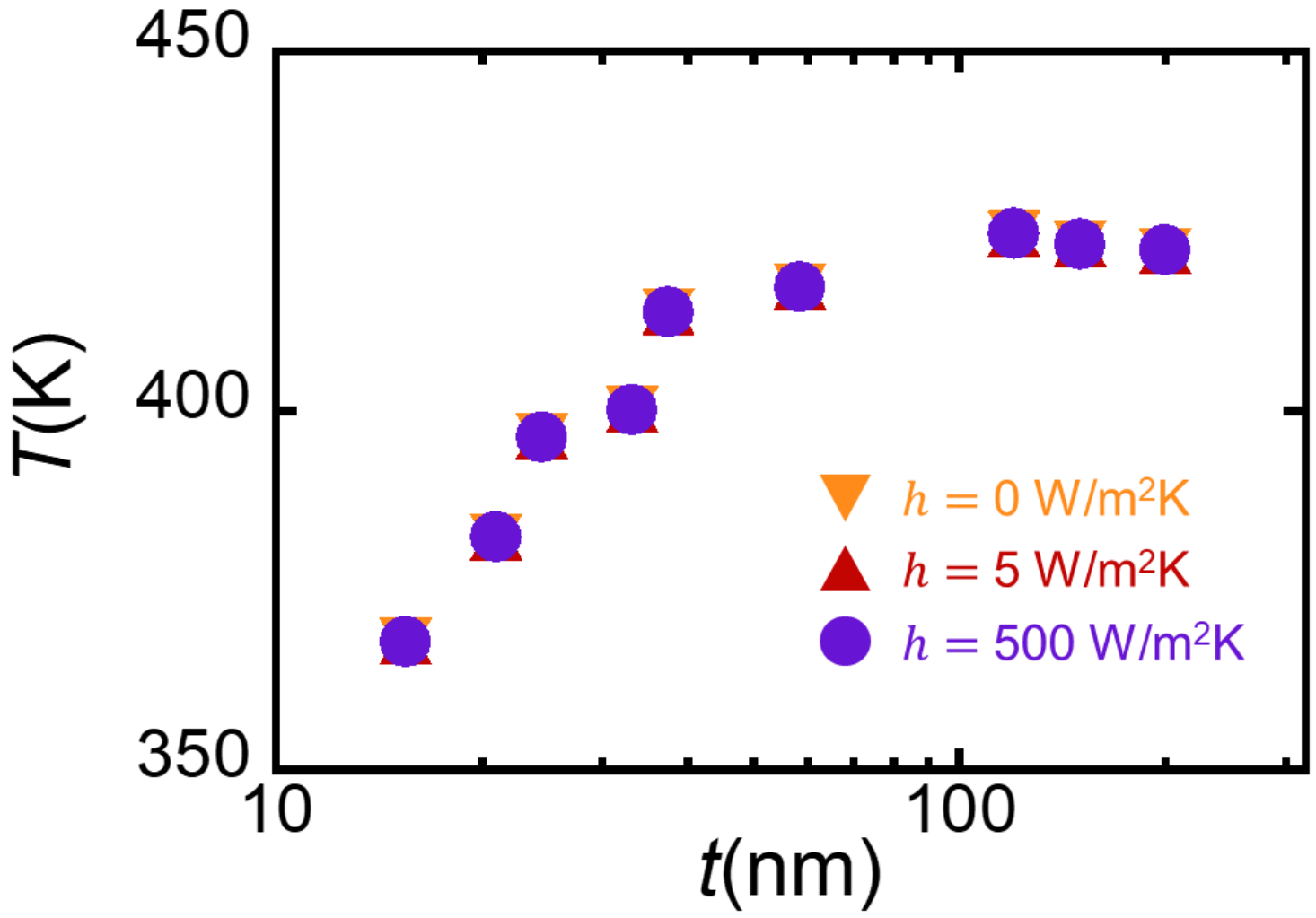

**Fig. S11.** Steady state temperatures for $Mn_3Sn(t)$/Ta(5 nm)/ devices on Si(380 μm)/$SiO_2$(500 nm) substrates under the observed critical current $I_C$ as a function of $t$ calculated using various heat transfer coefficients $h$. $h$ describes the convective heat flow into air used in the calculation. The results for $h = 5$ W/m$^2$K are identical to those shown in Fig. 5**c** of the main text.

**SM9: Thickness dependence of the switching current density in the heat-assisted switching mechanism**

Here we derive the thickness dependence of the current density required to heat a thin film device to a certain temperature, i.e., the switching current density in the heat-assisted switching mechanism. Consider a device with thickness $t$, channel width $w$, and length $l$ on a substrate, which in turn is placed on a sample stage that acts as a heat sink (Fig. S12**a**). Let $I$ be the current required to maintain a fixed device temperature (for instance, $T_N$). The Joule heating due to this current is given as

$$Q_{\mathrm{Joule}} = I^2 R$$

where $R$ is the resistance of the device. To make the thickness dependence explicit, we rewrite $I$ and $R$ as follows:

$$I = jwt, R = \frac{\rho l}{wt}$$

where $j$ is the current density and $\rho$ is the resistivity of the device. Plugging these expressions into $Q_{\mathrm{Joule}} = I^2 R$ gives

$$Q_{\text{Joule}} = (jwt)^2 \frac{\rho l}{wt} = j^2 wt\rho l.$$

On the other hand, we saw in SM8 that the heat transfer from the device to the surrounding environment is dominated by the heat flow into the substrate. This suggests that the environmental heat loss can be expressed as

$$Q_{\text{loss}} = \alpha S = \alpha wl$$

where $\alpha$ is a constant with respect to thickness and $S$ is the lateral area of the device.

The steady state condition $Q_{\text{Joule}} = Q_{\text{loss}}$ can thus be written as

$$j^2 wt\rho l = \alpha wl.$$

Solving for $j$ yields

$$j = \sqrt{\frac{\alpha}{\rho t}} \propto t^{-0.5}.$$

Note that the actual value of $j$ depends not only on film thickness but also on both the electrical resistivity ($\rho$) and the heat flow from the device to the substrate per unit contact area ($\alpha$). We check the validity of this relation in our device structure through numerical simulations. First, we simulate a monolayer $Mn_3Sn$ device with dimensions as shown in Fig. S10. In Fig. S12**b**, we show the values of $j$ that give a steady state device temperature of 420 K$\pm$0.1 K for $t$ ranging from 15 to 200 nm. The data is well described by the fit $j \propto t^{-0.5}$ (pink), consistent with our prediction.

Next, we investigate the effects of shunting into the Ta layer on the thickness dependence of $j$. We perform the same analysis described above for $Mn_3Sn$(*t*)/Ta(5) devices and again plot $j^{\text{Ta}}$ against $t$ in Fig. S12**c** ($j^{\text{Ta}}$ is the current density in the Ta layer when the device temperature saturates at 420 K$\pm$0.1 K). We see that $j^{\text{Ta}}$ scales as $t^{-0.5}$ at larger values of $t$, while the plots slightly deviate from this behavior in favor of smaller values of $j^{\text{Ta}}$ for smaller *t*. This deviation is understood by how the shunting effect becomes more prominent for thinner devices and is much more subtle than what we observe for thinner devices reported in Fig. 4**a** in the main text.

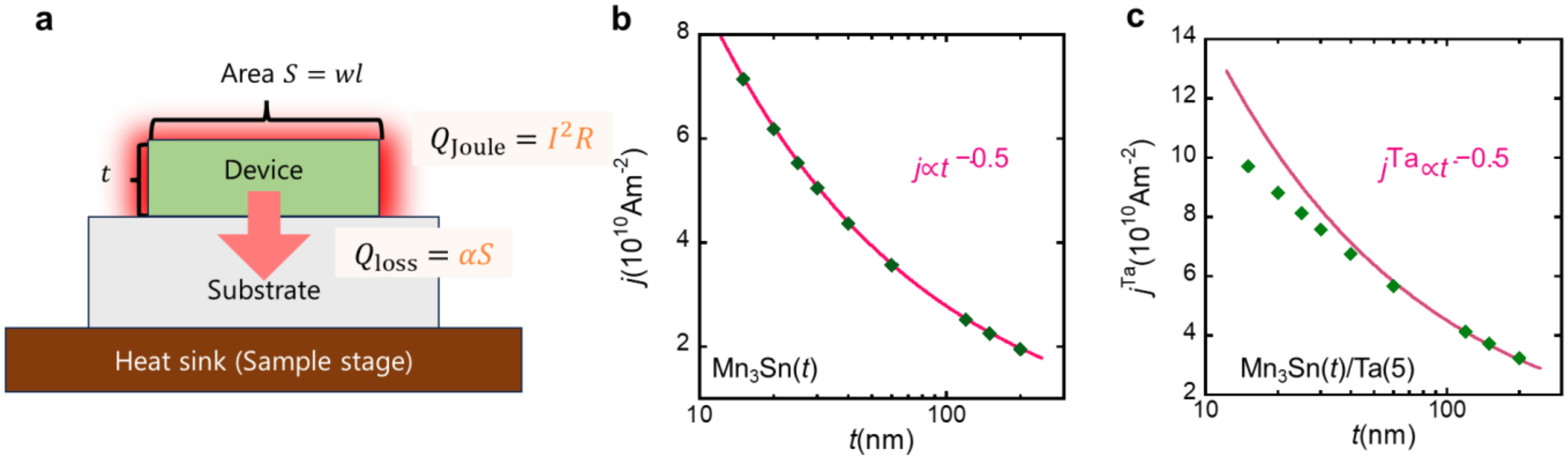

**Fig. S12. a**, Schematic of the relevant heat influx/outflux in the device. **b,** Calculated current density required to maintain a device temperature of 420 K (*j*) versus the thickness (*t*) of a $Mn_3Sn$ monolayer Hall bar on a Si/$SiO_2$(500 nm) substrate. **c,** Calculated current density in the Ta layer required to maintain a device temperature of 420 K ($j^{\mathrm{Ta}}$) versus the $Mn_3Sn$ layer thickness in a $Mn_3Sn$(*t*)/Ta(5 nm) bilayer Hall bar on a Si/$SiO_2$(500 nm) substrate. The pink lines indicate the fits for $j \propto t^{-0.5}$.

Finally, we comment on the thickness dependence of the power consumption required to keep the device at a fixed temperature. Our underlying assumption $Q_{\mathrm{Joule}} = Q_{\mathrm{loss}}$ (or $I^2R = \alpha S$) ensures that the power consumption involved in maintaining a constant device temperature through Joule heating is independent of thickness for a monolayer device. We examine the effect of the Ta layer on the thickness dependence of the power consumption in $Mn_3Sn$(*t*)/Ta(5) bilayer Hall bar devices by calculating the net power *P* as

$$P = \sum_i \rho^i (j^i)^2 v^i$$

where $i$ runs over the two materials in the bilayer ($Mn_3Sn$ and Ta) and $\rho^i, j^i$ and $v^i$ correspond to the resistivity, current density, and volumes of each layer. The values of $j^i$ are those calculated in the simulations with the $Mn_3Sn$(*t*)/Ta bilayer devices described above. The calculated thickness dependence of *P* is shown in Fig. S13. We see that shunting in the Ta layer has little effect on the net power consumed by the device in our entire thickness range. Therefore, when the temperature-assisted switching mechanism dominates, the switching power should be independent of thickness.

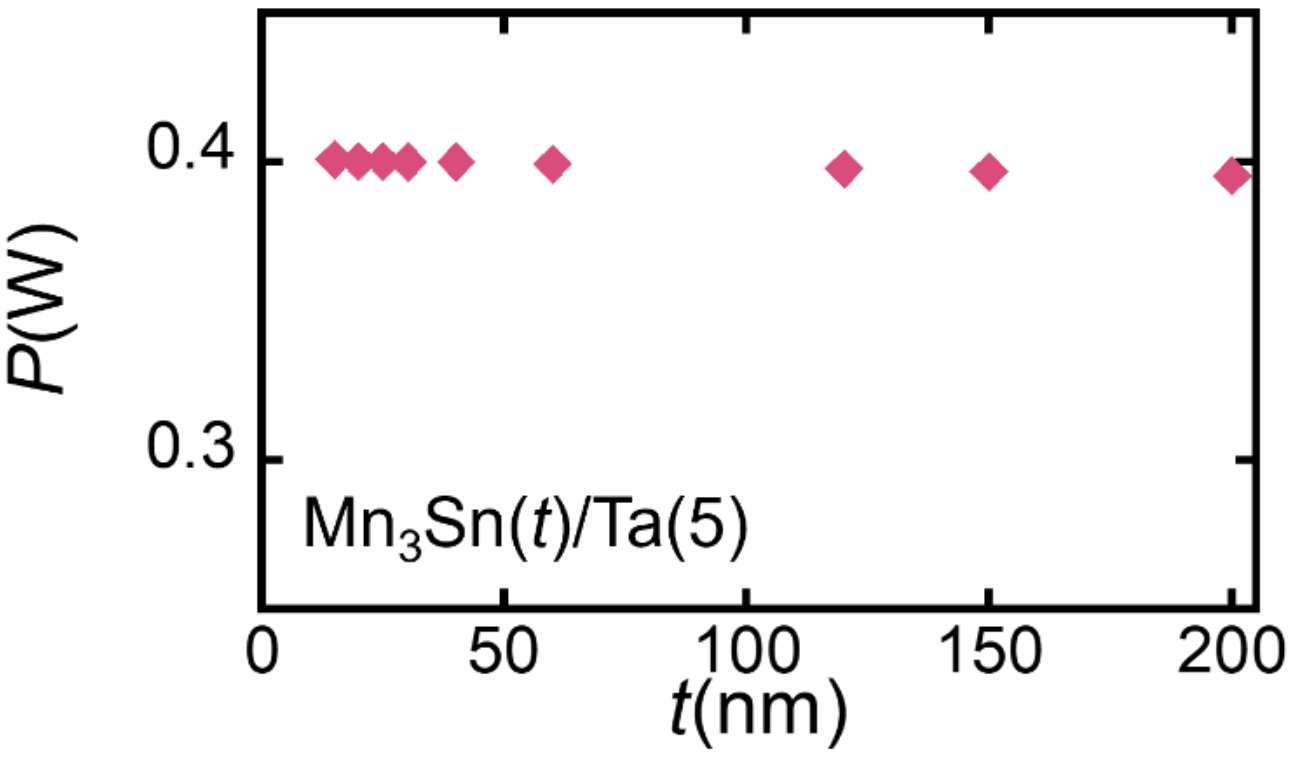

**Fig. S13.** Calculated power required to maintain a $Mn_3Sn$($t$)/Ta(5 nm) Hall bar on a Si/$SiO_2$(500 nm) substrate at a device temperature of 420 K.

**SM10. Time evolution of Joule heating**

To characterize this timescale associated with cooling in our devices, we have simulated the time evolution of Joule heating in $Mn_3Sn$($t$)/Ta(5 nm) devices under the experimentally observed critical current $I_C$. In the simulation, the current is turned on instantaneously at time $\tau = \tau_{start}$ =0 ns and is similarly instantaneously removed at $\tau = \tau_{end}$ =20000 ns; that is, the simulated pulse width is 20 μs, significantly longer than timescales reported for Joule heating in devices with similar structures and dimensions[12,13]. The simulated time evolution of the device temperature is shown in Fig. S14**a**, from which the characteristic time associated with cooling after the removal of current ($\tau_{cool}$) was estimated as the value that satisfies

$$T(\tau_{end} + \tau_{cool}) - T_0 = \frac{T_{steady} - T_0}{e},$$

Where $T_0$ is the temperature of the environment at 293.15 K, $T_{steady}$ is the steady state temperature under the current $I_C$ shown in Fig. 5**c** in the main text, and $e$ is Napier's constant. As shown in Fig. S14**b**, $\tau_{cool}$ increases with $t$ within the range of 200-600 ns for the thicknesses used in our study. This timescale matches the rise/fall time of the current pulses as well as the pulse widths for which suppression of heat-assisted switching was observed in previous works.

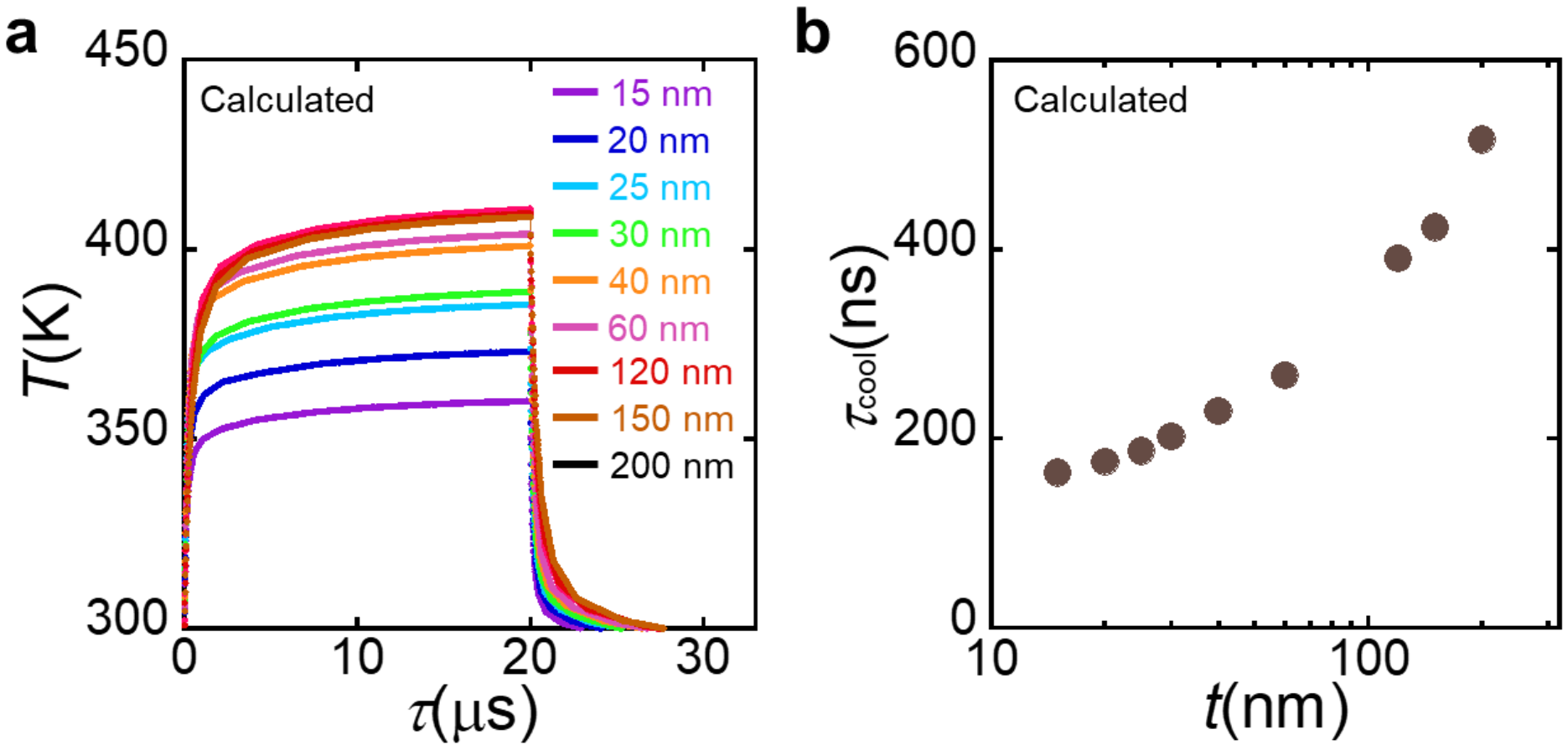

**Fig. S14. a**, Calculated time evolution of temperature under DC current $I_C$ in a $Mn_3Sn$($t$)/Ta (5 nm) Hall bar. The current is turned on at time $\tau = 0$ ns and removed at $\tau = 20000$ ns.) **b**, Thickness dependence of the simulated characteristic cooling relaxation time $\tau_{cool}$ in a $Mn_3Sn$($t$)/Ta (5 nm) Hall bar. The plotted values were calculated from the data shown in **a**.

**SM11. Relation between critical current density and switching volume**

In Fig. 4**c** of the main text, we saw that in our thicker films, i.e., $t>t_c$, the switching ratio $\xi$ rapidly decreases; in fact, $\xi$ is less than 2% at $t = 200$ nm. These values are significantly smaller than if the depth/volume of the $Mn_3Sn$ layer being switched is the same as those in the devices that satisfy $t \leq t_c$. For example, if the same depth/volume of $Mn_3Sn$ is switched in the $t$=30 nm and $t$=200 nm samples, we expect $\xi(t=200\ \text{nm})$ to be $\xi(t=30\ \text{nm}) \times \frac{30\ \text{nm}}{200\ \text{nm}} \approx 9\ \%$. This suggests that the switched $Mn_3Sn$ volume shrinks as the film thickness $t$ grows above $t_c$.

We find that critical current density in the Ta layer, $j_C^{Ta}$, is a good indicator of the switched volume of $Mn_3Sn$: $\xi t$, a quantity proportional to the $Mn_3Sn$ thickness switchable by current, is positively correlated with $j_C^{Ta}$, with a correlation coefficient of 0.81 (Fig. S15). Switching with smaller current density is a feature of thicker films (as indicated by the green plots in Fig. S15), where the switching is dominated by the thermal mechanism. Therefore, we conclude that suppressing the Joule heating associated with switching and enabling the application of large currents without triggering the heat-assisted switching mechanism can enhance the volume of $Mn_3Sn$ that can be switched by SOT.

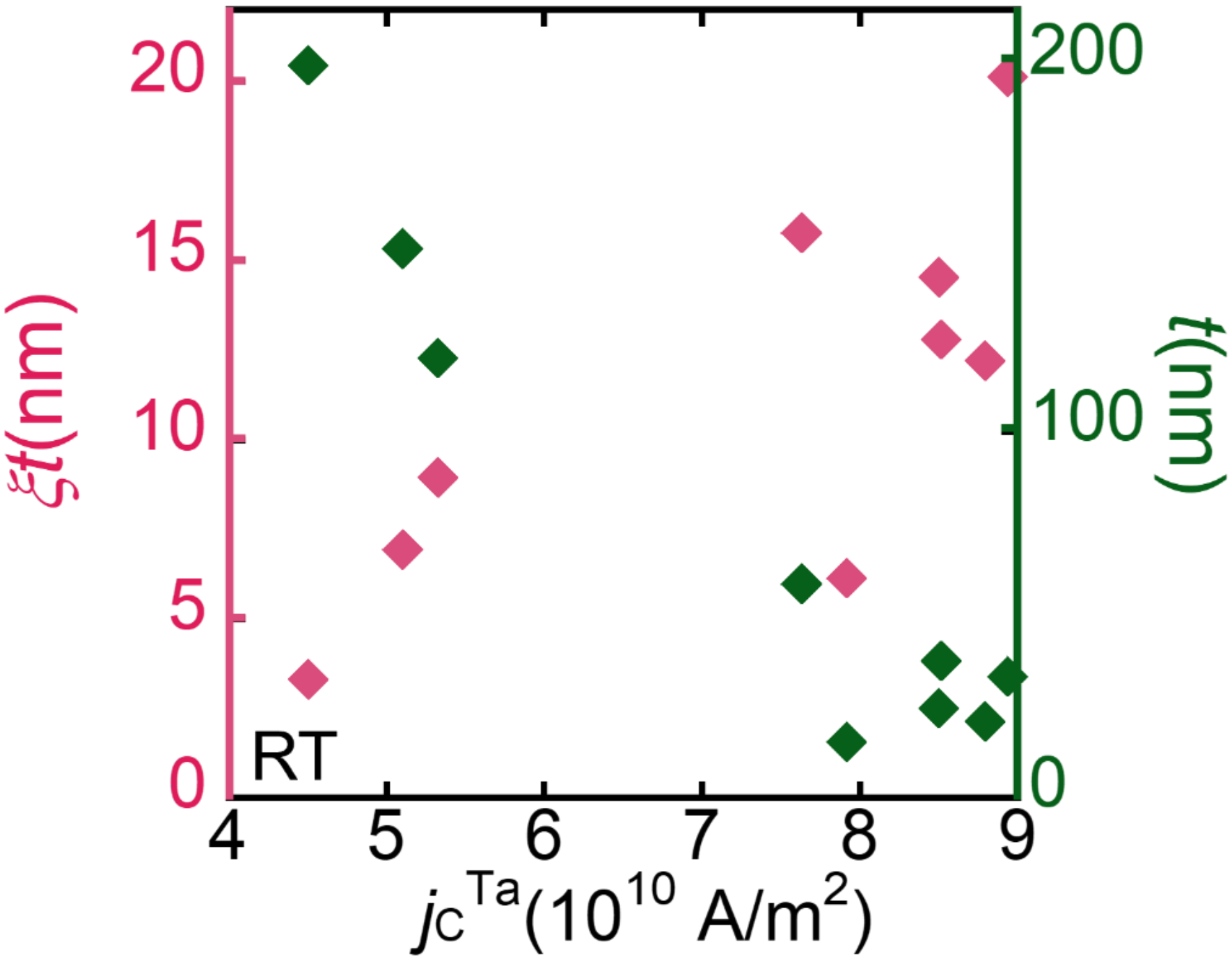

**Fig. S15.** $\xi t$ (an indicator of the thickness of the switched $Mn_3Sn$ region) (pink) plotted against $j_C^{Ta}$. The correlation coefficient between the two quantities was calculated to be 0.81, indicating a large positive correlation. The relationship between $j_C^{Ta}$ and $t$ (green) is also shown to highlight that $\xi t$ shrinks with growing $t$.

**SM12. Determining the transition temperature through the anomalous Hall effect**

Knowledge of the transition temperature $T_{\mathrm{trans}}$ for our films is required to determine the relation between the switching temperature and the destabilization of the antiferromagnetic structure. The transition temperatures for ferromagnetic films have been shown to decrease as they become thinner[15,16], and a similar effect has been observed in $Mn_3Sn$[17]. Therefore, the temperature dependence of the anomalous Hall loops induced by a perpendicular magnetic field $H$ in our $Mn_3Sn(t)$/Ta(5 nm)/$AlO_x$(3 nm) Hall bars was investigated to characterize any systematic dependence of $T_{\mathrm{trans}}$ on film thickness. The temperature was controlled by the sample stage temperature $T_{\mathrm{stage}}$, and the current used to probe the AHE in these measurements was set to be less than 2 % of the critical currents for switching ($I_{\mathrm{C}}$) in each device measured at room temperature to minimize Joule heating. Under these conditions, the sample temperature can be well approximated by $T_{\mathrm{stage}}$.

The $T_{\mathrm{stage}}$ dependence of the spontaneous Hall resistance $\Delta R_{\mathrm{H}}^{\mathrm{field}}$ (defined as $R_{\mathrm{H}}(\mu_0 H = -0) - R_{\mathrm{H}}(\mu_0 H = +0)$, where $\mu_0 H = +0$ indicates $\mu_0 H = 0$ T while sweeping from positive to negative and vice versa) for each sample, normalized by their values at $T_{\mathrm{stage}} =$ 300 K, are shown in Fig. S16. $T_{\mathrm{trans}}$ was estimated as the value of $T_{\mathrm{stage}}$ at which $\Delta R_{\mathrm{H}}^{\mathrm{field}}$ becomes 1% of that at 300 K. The obtained values for $T_{\mathrm{trans}}$ are shown in Fig. 5**b** of the main text.

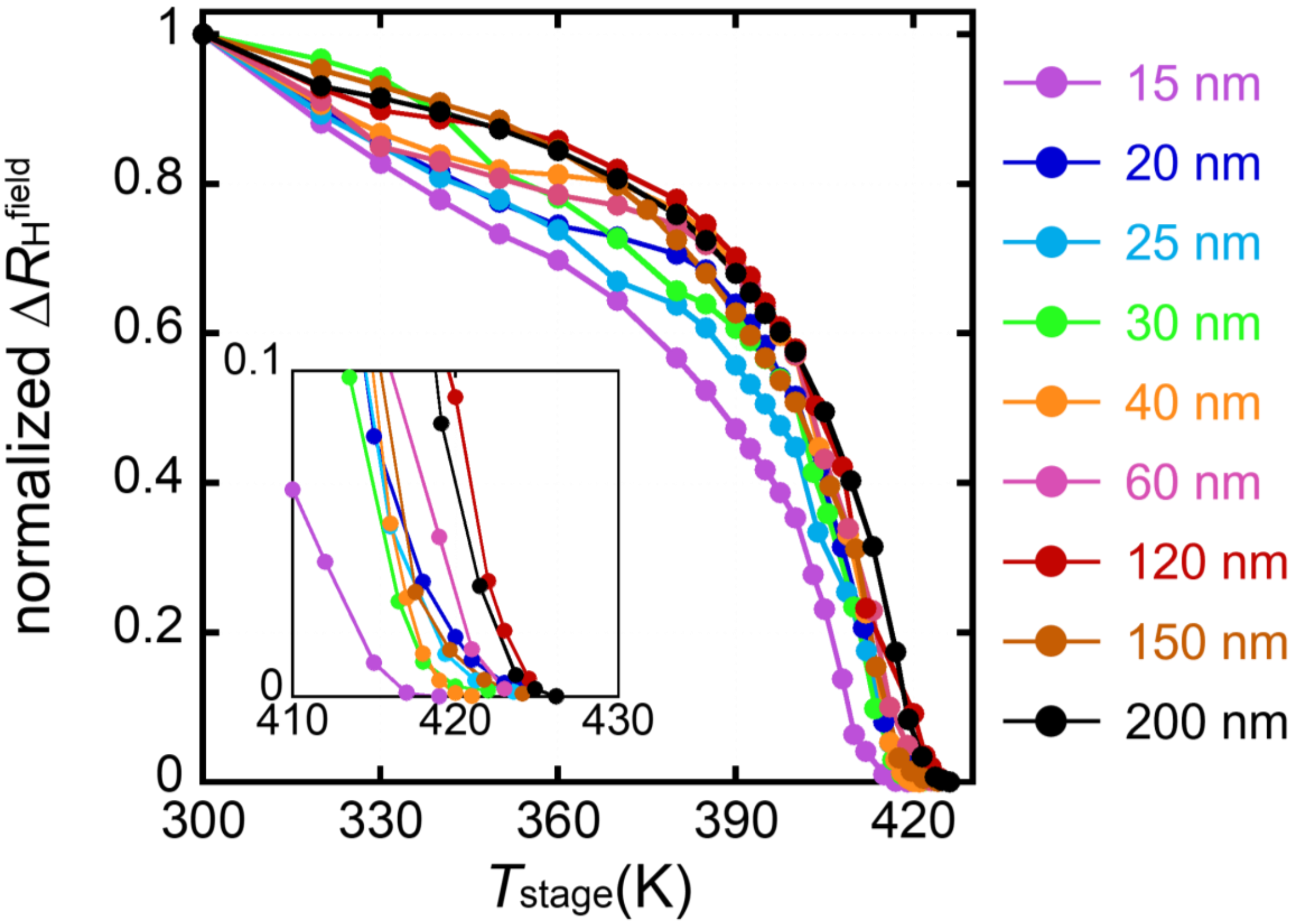

**Fig. S16.** Evolution of the spontaneous anomalous Hall effect as a function of sample stage temperature $T_{\mathrm{stage}}$ measured for the Hall bars fabricated from $Mn_3Sn$($t$)/Ta(5 nm)/$AlO_x$(3 nm) films on Si(380 μm)/$SiO_2$(500 nm) substrates. The values of the Hall resistance on the $y$-axis are normalized by their values at $T_{\mathrm{stage}} = 300$ K. Each curve represents the values measured for a particular $t$, as indicated in the legend. The inset is a magnification of the data

around the observed transition temperatures.

## SM13. $\Delta R_{\mathrm{H}}^{\mathrm{elec}}$ as a function of $H_{\mathrm{bias}}$

We plot the switching signal $\Delta R_{\mathrm{H}}^{\mathrm{elec}}$ as a function of bias field in Fig. S17. In all cases, $|\Delta R_{\mathrm{H}}^{\mathrm{elec}}|$ increases with $H_{\mathrm{bias}}$, peaks at a certain field value, then starts to approach zero at larger bias fields. While several previous works have suggested that this nonmonotonic behavior as evidence for the temperature-assisted mechanism[12,13], we argue that it can be explained regardless of the switching mechanism. (We also point out that the reports on the temperature-assisted mechanism only point out that this behavior does not match reports in conventional perpendicular ferromagnets-neither provide an explanation for why this can be interpreted as a unique fingerprint of the temperature-assisted mechanism.)

The nonmonotonic dependence of $\Delta R_{\mathrm{H}}^{\mathrm{elec}}$ on $H_{\mathrm{bias}}$ can be interpreted as a result of the antiferromagnetic order (or more accurately, the weak ferromagnetic moment associated with the order) being forced to lie along the bias field through Zeeman coupling. The discrepancy between this behavior in $Mn_3Sn$ and conventional perpendicular ferromagnets can be explained by how the strong PMA in ferromagnetic devices make it difficult to orient the magnetization along the in-plane bias field, while the magnetic order in polycrystalline or textured $Mn_3Sn$ (which is the focus of these reports) does not have a strong easy axis with respect to the device geometry and is therefore more susceptible to reorientation by the field[18]. That being said, a similar suppression of the switching signal at large bias fields, while not as dramatic as the $Mn_3Sn$ case, has been seen even for ferromagnetic CoFeB with a large PMA[12]. (Ref.[12], which reports the bias field dependence of the SOT switching signal of CoFeB, describe it as “almost independent” compared to the $Mn_3Sn$ case, the latter of which is the main focus of their study. However, this difference can be explained simply by the difference in the magnetic anisotropy in the two materials. It should also be noted that the largest bias field they employ for CoFeB switching is an order of magnitude smaller than what they use for $Mn_3Sn$.) The nonmonotonic dependence of the switching signal can therefore be understood as a manifestation of the competition of the Zeeman coupling described above and the efficiency of the damping-like torque that is supplemented by the bias field.

One noticeable difference in the $\Delta R_{\mathrm{H}}^{\mathrm{elec}}$-$H_{\mathrm{bias}}$ curves between the intrinsic mechanism ($t$=20 nm) (i.e., the sample temperature during SOT-induced switching remains well below $T_{\mathrm{N}}$) and the temperature-assisted mechanism ($t$=60, 120, 200 nm) (i.e., the sample temperature during switching rises close to $T_{\mathrm{N}}$) is the value of $H_{\mathrm{bias}}$ required to suppress the switching signal. For the $t$=20 nm device, $\Delta R_{\mathrm{H}}^{\mathrm{elec}}$ peaks around 0.05 to 0.2 T and starts to significantly decrease only at $|\mu_0 H_{\mathrm{bias}}| \geq 0.3$ T, while $\Delta R_{\mathrm{H}}^{\mathrm{elec}}$ disappears much more rapidly in response to the in-plane $H_{\mathrm{bias}}$. We hypothesize that this increased susceptibility to the in-plane bias field for the temperature-assisted mechanism is due to how the actual SOT-induced switching process happens at temperatures just below the transition temperature $T_{\mathrm{trans}}$, where the magnetism is as soft as it can be. In the intrinsic mechanism, the magnetic structure is intact with a finite magnetic anisotropy, and as such a larger magnetic field is required to align the magnetic order in the direction of the bias field during all phases of the switching process compared to the temperature-assisted switching process. In other words, the change in the bias field dependence should represent the difference in the switching mechanism.

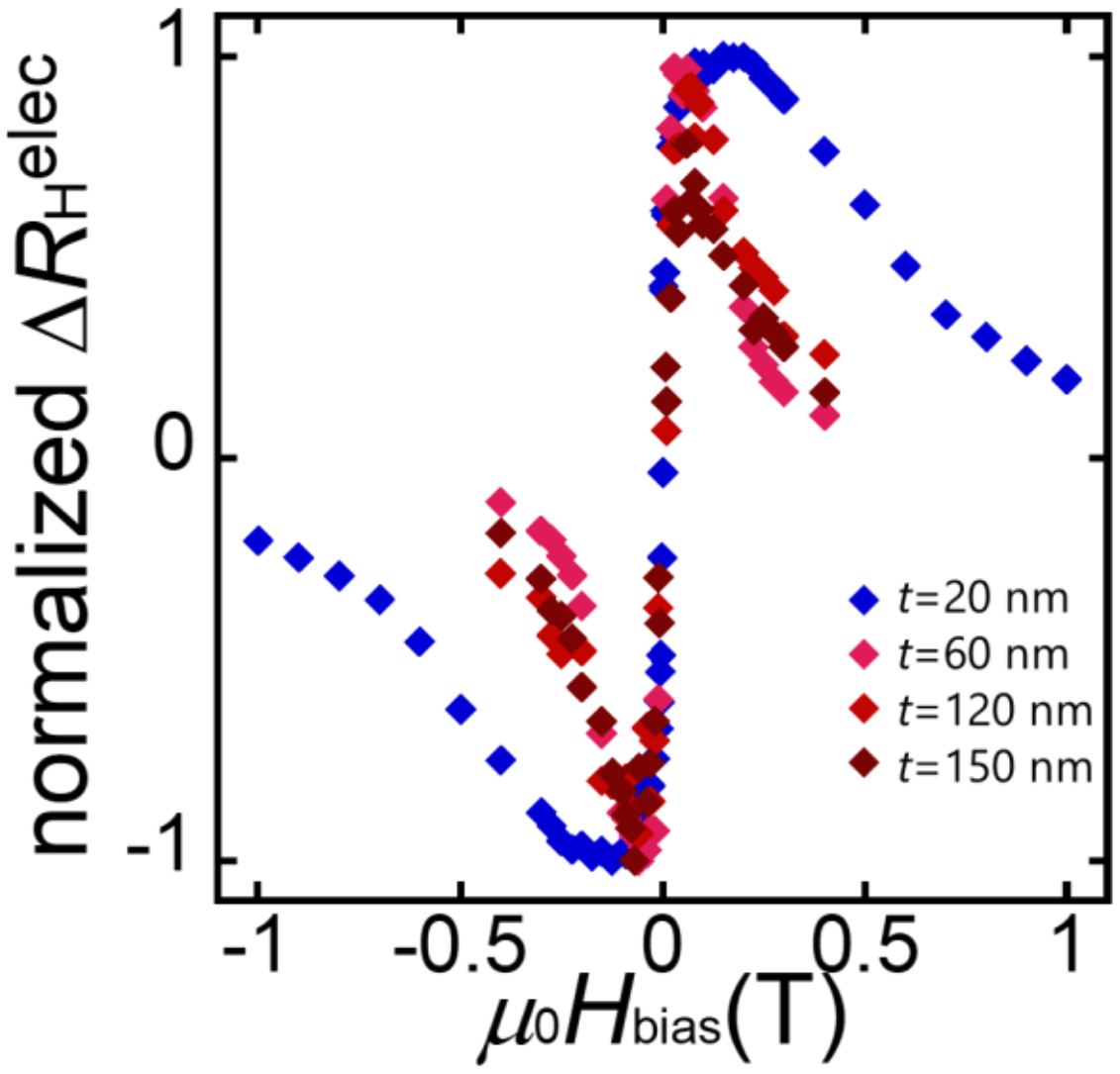

**Fig. S17.** Switching signal as a function of bias field for devices with $t$ ranging from 20 nm to 150 nm. The values are normalized by the maximum values in each dataset for comparison across different thicknesses.

## SM14. Switching using 100 ms single-shot pulses

To implement $Mn_3Sn$ into magnetic memory architecture, the antiferromagnetic order must be switchable with a single-shot pulse. In Fig. S18, we show the Hall resistance after applying a write pulse $I_{\mathrm{write}}$ ($\tau_{\mathrm{pulse}}$ = 100 ms) of fixed magnitude (>$I_{\mathrm{C}}$) and varying polarity for devices with $t$ = 20 nm (representative of the intrinsic mechanism) and 150 nm (representative of the temperature-assisted mechanism) measured by a read current of 0.2 mA. We see that in both cases, the switching behavior is dictated based on the polarity of the pulse, and that the magnetic state is stable against consecutive injections of pulses with the same polarity. Importantly, the switching signal observed in these single-shot experiments are consistent with those in the switching loops where multiple pulses were applied before the device was fully switched, showing that these single-shot pulses fully switch the device.

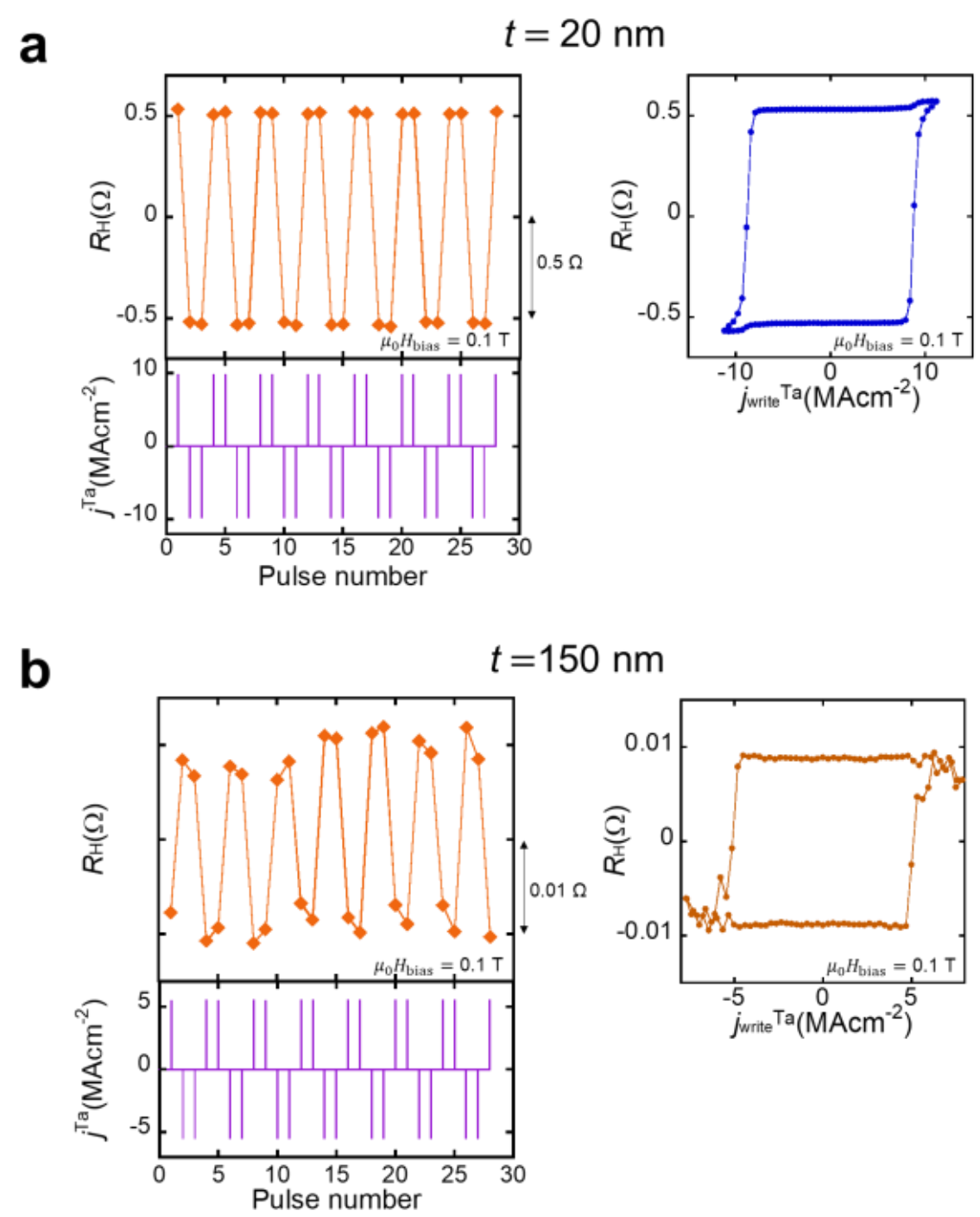

**Fig. S18.** Switching achieved with current pulses of fixed magnitudes for devices with $t$=20 nm (**a**) and $t$=150 nm (**b**). The sign of the Hall resistance is determined solely by the polarity of the pulse injected prior to measurement. The right panels of each figure show the switching loops obtained by injecting successive write pulses while sweeping the amplitude.

**SM15.** Robustness of thickness-driven SOT switching mechanism crossover against definition of switching current

In this study, we have defined the switching current $I_{\mathrm{C}}$ as the write current at which the Hall signal changes sign. However, here we show that our conclusions of a crossover in the switching mechanism across a $Mn_3Sn$ thickness $t_{\mathrm{c}}$ remains unchanged regardless of whether we analyze the current at which switching is just barely triggered, the current at which half of the crystal grains in the Hall device are switched ($I_{\mathrm{C}}$), or the current at which most of the crystal grains can be switched.

Let us define $I(n\%)$ as the write pulse current amplitude that switches $n\%$ of the total switchable region in the Hall bar. From our switching loops, we can calculate $I(n\%)$ as the current value that satisfies $\frac{|R_{\mathrm{H}}(I_{\mathrm{write}}=I(n\%))-R_{H}(I_{\mathrm{write}}=0)|}{\Delta R_{\mathrm{H}}^{\mathrm{elec}}}=n\%$ where $R_{\mathrm{H}}(I_{\mathrm{write}}=I(n\%))$ refers to the Hall resistance measured with $I_{\mathrm{read}}=0.2$ mA after the application of a write pulse of duration 100 ms with amplitude $I_{\mathrm{write}}=I(n\%)$. Note that $I(50\%)=I_{\mathrm{C}}$. Similarly, we define $j^{\mathrm{Ta}}(n\%)$ as the current density in the Ta layer when current $I(n\%)$ is flowing in the device, $V(n\%)$ as the voltage applied to the device to generate a current of $I(n\%)$, and $P(n\%)=I(n\%)V(n\%)$ as the dissipated power in the device when current $I(n\%)$ is flowing.

When we plot $j^{\mathrm{Ta}}(n\%)$ and $P(n\%)$ as a function of $Mn_3Sn$ thickness for $n=20$, 50, and 80 (Fig. S19**a**), we see for all $n$ the crossover in behavior we saw for $j_{\mathrm{C}}^{\mathrm{Ta}}(=j^{\mathrm{Ta}}(50\%))$ and $P_{\mathrm{C}}(=P(n\%))$ across the crossover thickness $t_{\mathrm{c}}\approx 30$ nm. Specifically, $j^{\mathrm{Ta}}(n\%)$ generally follows $t^{-0.5}$ in the thickness range $t>t_c$, and deviates to smaller values at $t<t_c$, consistent with our prediction shown in Fig. 1**c** of the main text. Meanwhile, $P(n\%)$ is constant in the range $t>t_c$ (Fig. S19**b**), consistent with Ref. [29] in the main text which reports on temperature-assisted switching of $Mn_3Sn$, and decreases when $t<t_c$, indicating a suppression in the switching temperature as the thickness is reduced below $t_c$.

Finally, we estimate the device temperatures $T(n\%)$ ($n=20$, 50, and 80) for each $t$ when $I(n\%)$ is applied with the same methodology employed in the main text for when $I_{\mathrm{C}}(=I(50\%))$ is applied and compare them with the transition temperatures $T_{\mathrm{trans}}$ (Fig. S19**c**). Consistent with the behavior of $P(n\%)$ as a function of $t$, $T(n\%)$ falls below $T_{\mathrm{trans}}$ when $t$ is reduced below $t_c$. This behavior again mimics the case of $T^*(=T(50\%))$ presented in Fig. 5**b**.

Based on these results, we conclude that our definition of $I_C$, corresponding to $I(50\%)$, is appropriately representative of the switching properties of our devices.

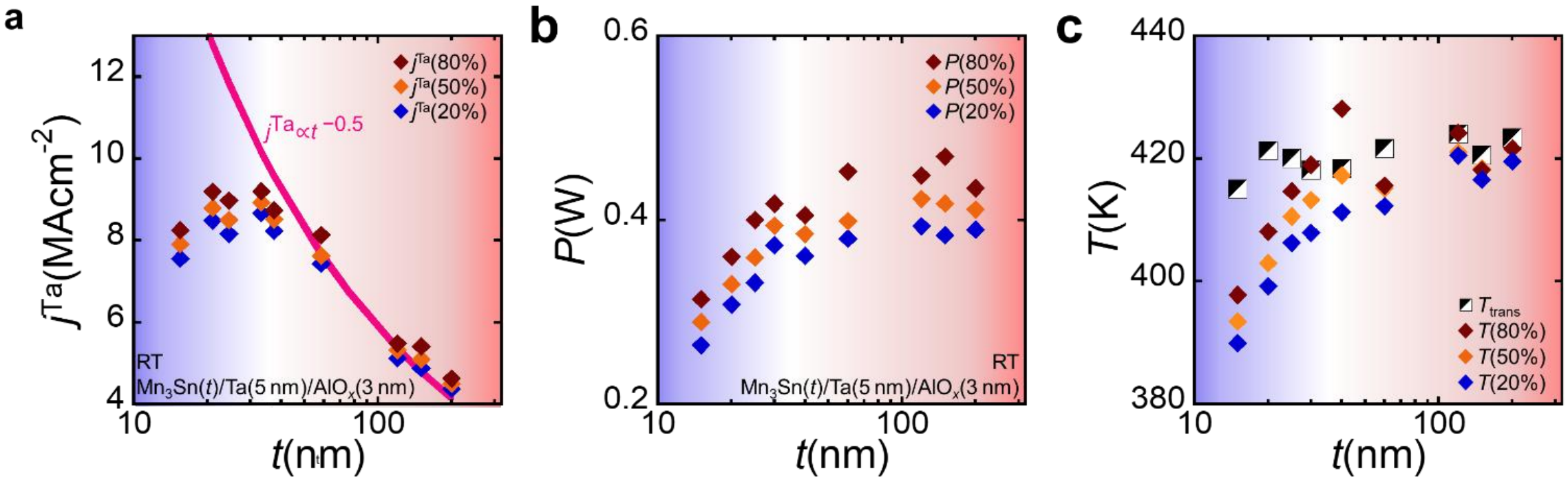

**Fig. S19.** **a**, Current density in the Ta layer required to achieve a $n$% switching amplitude ($j^{\mathrm{Ta}}(n\%)$), where $n$ =20, 50, and 80. **b**, Power associated with a $n$% switching amplitude. **c**, Steady state device temperature when $I(n)$% ($n$ =20, 50, and 80.) is applied to the device compared to the transition temperature $T_{\mathrm{trans}}$.

**References in Supplementary Material**

1. Higo, T. *et al.* Anomalous Hall effect in thin films of the Weyl antiferromagnet $Mn_3Sn$. *Appl. Phys. Lett.* **113**, 202402 (2018).

2. Ikeda, T. *et al.* Improvement of Large Anomalous Hall Effect in Polycrystalline Antiferromagnetic $Mn_{3+x}Sn$ Thin Films. *IEEE Trans. Magn.* **55**, 1–4 (2019).

3. Duan, T. F. *et al.* Magnetic anisotropy of single-crystalline $Mn_3Sn$ in triangular and helix-phase states. *Appl. Phys. Lett.* **107**, 082403 (2015).

4. Sung, N. H., Ronning, F., Thompson, J. D. & Bauer, E. D. Magnetic phase dependence of the anomalous Hall effect in $Mn_3Sn$ single crystals. *Appl. Phys. Lett.* **112**, 132406 (2018).

5. Yoon, J.-Y. *et al.* Correlation of anomalous Hall effect with structural parameters and magnetic ordering in $Mn_{3+x}Sn_{1-x}$thin films. *AIP Adv.* **11**, 065318 (2021).

6. Nakatsuji, S., Kiyohara, N. & Higo, T. Large anomalous Hall effect in a non-collinear antiferromagnet at room temperature. *Nature* **527**, 212–215 (2015).

7. Higo, T. *et al.* Perpendicular full switching of chiral antiferromagnetic order by current. *Nature* **607**, 474–479 (2022).

8. Tomita, T., Ikhlas, M. & Nakatsuji, S. Large Nernst Effect and Thermodynamics

Properties in Weyl Antiferromagnet. in *Proceedings of the International Conference on Strongly Correlated Electron Systems (SCES2019)* (Journal of the Physical Society of Japan, Okayama, Japan, 2020). doi:10.7566/JPSCP.30.011009.

9. Online Materials Information Resource - MatWeb. https://www.matweb.com/.
10. Cook, H. C. Investigation of Sputtered β-Tantalum Thin Films. *J. Vac. Sci. Technol.* **4**, 80–86 (1967).
11. *Heat Transfer Handbook*. (Wiley, Hoboken, NJ, 2003).
12. Pal, B. *et al.* Setting of the magnetic structure of chiral kagome antiferromagnets by a seeded spin-orbit torque. *Sci. Adv.* **8**, eabo5930 (2022).
13. Krishnaswamy, G. K. *et al.* Time-Dependent Multistate Switching of Topological Antiferromagnetic Order in $Mn_3Sn$. *Phys. Rev. Appl.* **18**, 024064 (2022).
14. Materials Data on $Mn_3Sn$ by Materials Project. LBNL Materials Project; Lawrence Berkeley National Laboratory (LBNL), Berkeley, CA (United States) https://doi.org/10.17188/1197606 (2020).
15. Yamaguchi, A. *et al.* Effect of Joule heating in current-driven domain wall motion. *Appl. Phys. Lett.* **86**, 012511 (2005).
16. Huang, F., Mankey, G. J., Kief, M. T. & Willis, R. F. Finite-size scaling behavior of

ferromagnetic thin films. *J. Appl. Phys.* **73**, 6760–6762 (1993).

17. Takeuchi, Y. *et al.* Chiral-spin rotation of non-collinear antiferromagnet by spin–orbit torque. *Nat. Mater.* **20**, 1364–1370 (2021).

18. Higo, T. *et al.* Omnidirectional Control of Large Electrical Output in a Topological Antiferromagnet. *Adv. Funct. Mater.* **31**, 2008971 (2021).